\documentclass[aps,prb,twocolumn,groupedaddress]{revtex4}

\usepackage{amsmath}
\usepackage{amsfonts}
\usepackage{siunitx}
\usepackage{xcolor}
\usepackage{graphicx}
\usepackage{bm}
\usepackage[caption=false]{subfig}
\graphicspath{ {./Images/} }

\newcommand{\ket}[1]{|#1\rangle}
\newcommand{\bra}[1]{\langle #1|}

\begin{document}

\title{Correlated physics in an artificial triangular anti-dot lattice}

\author{Z. E. Krix}
\email[]{z.krix@unsw.edu.au}
\author{H. D. Scammell}
\author{O. P. Sushkov}
\affiliation{School of Physics UNSW, Sydney}

\date{\today}

\begin{abstract}
This work considers a two-dimensional artificial triangular anti-dot lattice (TAL); a semiconductor based artificial crystal hosting Dirac cones, flat bands and Fermi surface nesting. All such single particle features have dramatic implications for the emergent correlated phases. This work predominantly focuses on the existence of a robust flatband and enumerates the possible correlated phases that follow. We find that the flatband is generated, in the single-particle theory, when charges align themselves along a kagome lattice with the same period as the TAL. The correlated phases are studied using complementary techniques of expansions in strong and weak Coulomb interaction. Our microscopic modelling shows that for the purpose of generating strongly correlated phases, hole doped TALs have significant advantages over electron doped.
\end{abstract}

\maketitle

\section{Introduction}

Systems hosting flat bands have been a strong focus of recent theoretical and experimental investigations. Moire superlattice structures such as twisted bilayer graphene\cite{balents_superconductivity_2020} and twisted transition metal dichalcogenides\cite{zhang_flat_2020} as well as kagome metals\cite{kang_dirac_2020} all host flat bands. Interest in these systems stems from the enhanced effect of Coulomb interactions within the flat band. A striking example of this is the emergence of superconductivity in twisted bilayer graphene\cite{cao_unconventional_2018}. In addition to superconductivity, ferromagnetic and charge density wave phases have been predicted\cite{liu_exotic_2014}. Particularly relevant and exciting experimental results come from the Vanadium family of kagome metals, which have recently been found to host superconducting and charge-density-wave ground states \cite{Ortiz2020,Zhu2021,Chen2021,Ortiz2021,Ni2021,Jiang2021}.

%$\text{AV}_{3} \text{Sb}_{3}$ ($\text{A} = \text{K}$, Rb,Cs

In this work we consider a flat band which can be generated in ordinary semiconductors via periodic electrostatic gating. When the applied potential has hexagonal symmetry and is sufficiently repulsive, a 2D system of electrons or holes will develop a flat band and two pairs of Dirac cones\cite{tkachenkoEffectsCoulombScreening2015}. An advantage of this approach to generating a flatband is the ability to tune its band width by varying either the modulation strength or the lattice period (which can be tens to hundreds of nanometres). In what follows we refer to these systems as triangular anti-dot lattices (TAL) and, for concreteness, we consider electrons and holes in GaAs. An experimental realisation of the TAL in the weak to moderate modulation regime has been developed and the results are to be published\cite{wangUnpublished2021}. Excitingly, there has been recent experimental progress in the electron based TAL \cite{Pinczuk2021PRL}, which demonstrated key band structure features. Although it has not yet considered the effective kagome bands discussed here, nor the use of holes (instead of electrons). Here we show that the hole band structure (which must account for spin-orbit coupling) has the same features of interest, i.e. a flat band and Dirac points and we present a numerical technique for computing this.

We have found that the flat band of TALs, in a system with either holes or electrons, can be described by an effective tight-binding model on an emergent kagome lattice. We compute the effective on-site Hubbard energy, $U_{0}$, and hopping parameter, $t$, and we show that $U_{0} / t \lesssim 10$ for electrons and $ \gtrsim 20$ for holes. Within this effective model we consider a number of correlated phases. Specifically, we consider commensurate and incommensurate charge density waves (CDWs), the Mott insulator phase, Stoner ferromagnetism and electron-electron driven superconductivity. The effective kagome model has significant consequences for the CDW and Mott phases since this affects the geometry of the CDW patterns and the particle densities at which they occur. We present a set of CDW patterns on the kagome lattice for different filling fractions of the flat band and identify which patterns minimise the Coulomb energy. Using our band structure calculations, we show that a TAL with holes develops a flat band at a much weaker modulation than electrons and has significantly stronger interactions.

The electronic band structure and mapping to a Hubbard model is discussed in Section \ref{sec:hubbard}. Section \ref{sec:holes} covers the hole band structure and compares the electron and hole flat bands. Sections \ref{sec:strong} and \ref{sec:weak}, discuss possible strongly and weakly correlated phases.

\begin{figure*}[t!]
\centering
\subfloat
    [\label{fig:AGbandstructure_W=1p0}]
    {\includegraphics[width=0.33\textwidth]{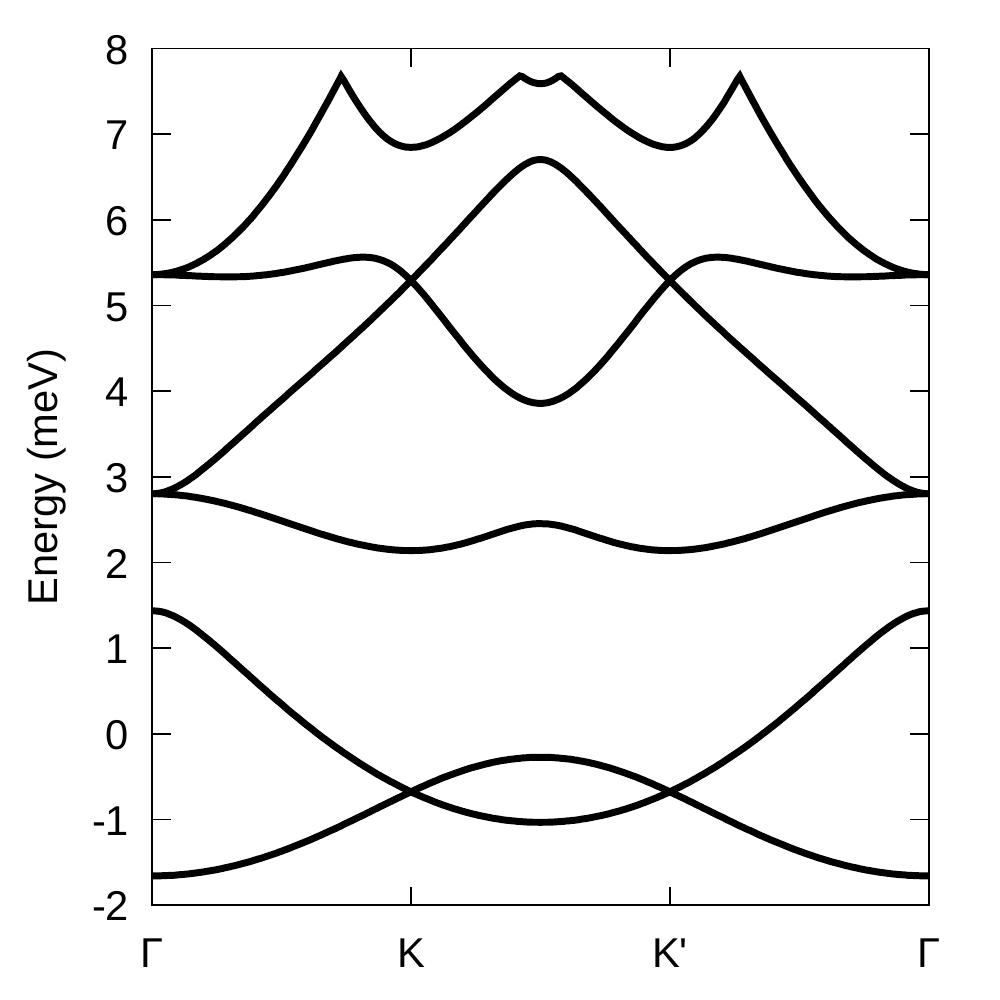}}
\subfloat
    [\label{fig:AGbandstructure_W=2p5}]
    {\includegraphics[width=0.33\textwidth]{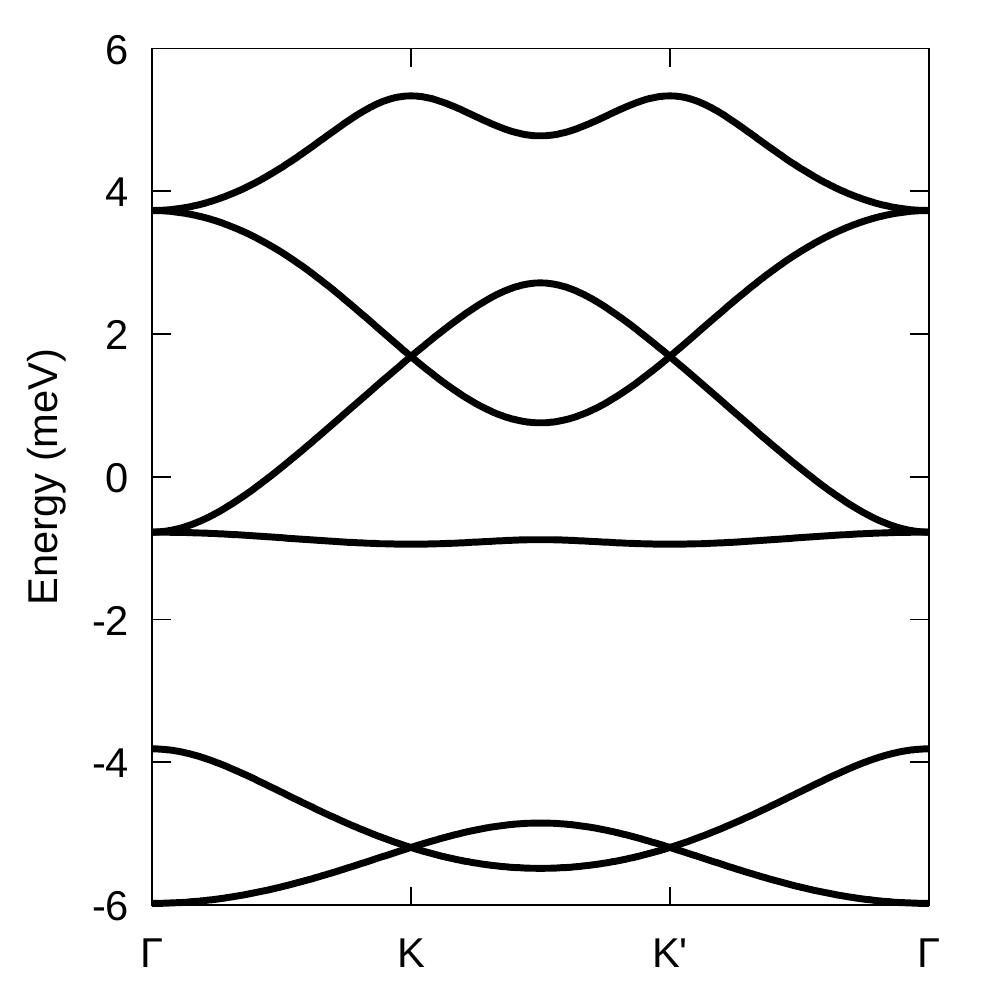}}
\subfloat
    [\label{fig:kagomebandstructure}]
    {\includegraphics[width=0.33\textwidth]{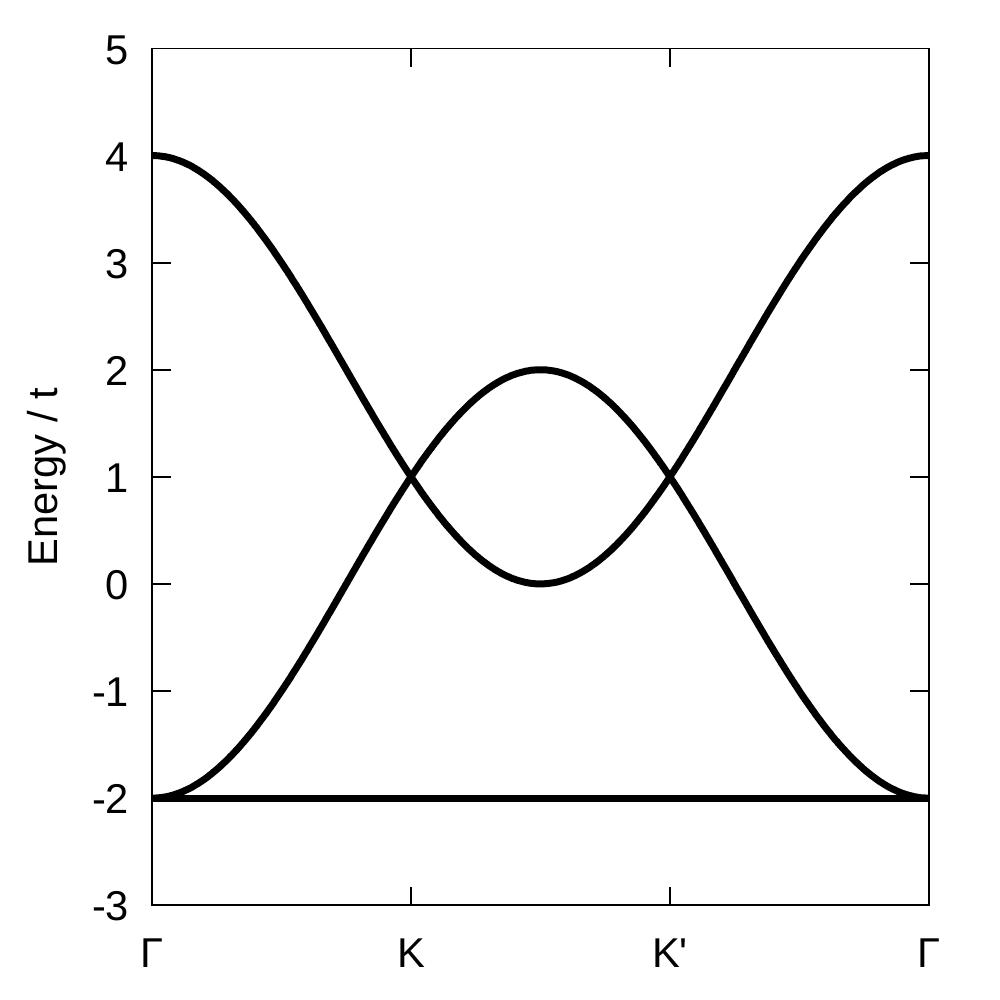}}

\subfloat
    [\label{fig:BZ}]
    {\includegraphics[width=0.25\textwidth]{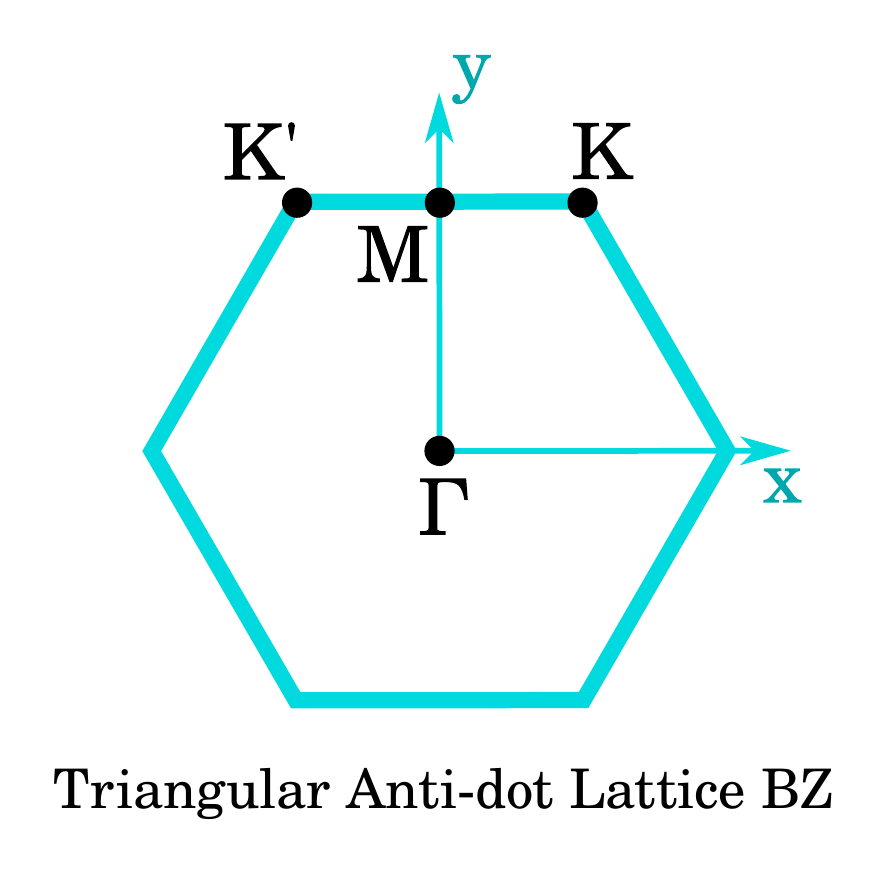}}
\subfloat
    [\label{fig:AG_FB}]
    {\includegraphics[width=0.4\textwidth]{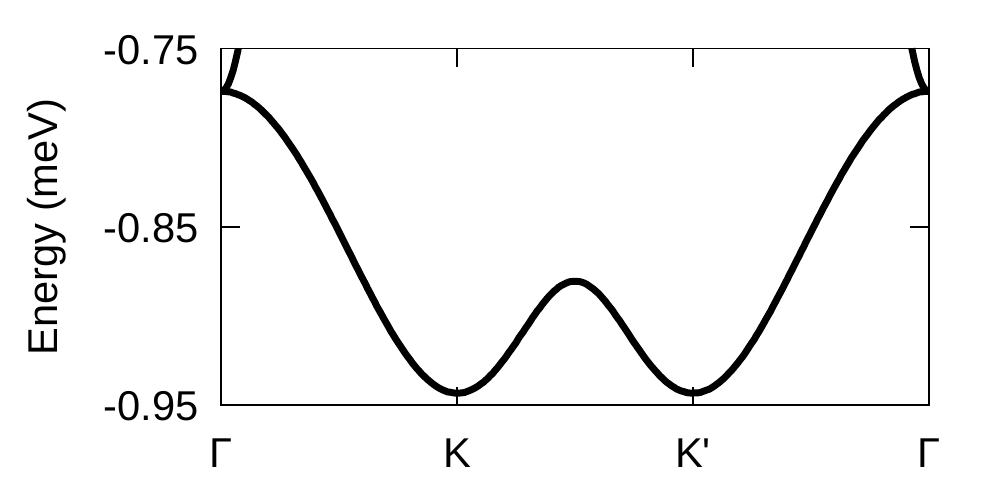}}

\caption{(a-c) Left: First six bands of the TAL electronic band structure. The lattice constant is $a = 80$ nm and $W = 1.0 E_{0}$ with $E_{0} = 1.57$ meV. Here, $\Gamma$ is the origin and $K$ and $K'$ are the two non-equivalent Brillouin zone vertices. Middle: The same band structure for $W = 2.5 E_{0}$. Right: The three energy bands of the kagome lattice with nearest neighbour hopping $t > 0$. (d) Schematic of the Brillouin zone for the triangular lattice represented by Eqn. \ref{equ:latticePotential}. (e) Close up of the flat band in Fig. \ref{fig:AGbandstructure_W=2p5}. While there is non-zero band curvature here, it is small compared with the total width of the three kagome-like bands in Fig. \ref{fig:AGbandstructure_W=2p5}.}
\label{fig:AGplusKagome_bandstructure}
\end{figure*}

\section{Electron flat band and effective kagome model}\label{sec:hubbard}

Within a single-electron model we are able to compute both the energy levels and eigenfunctions of the TAL Hamiltonian $H = \bm{p}^{2} / 2m + U(\bm{r})$ by numerical diagonalisation. The superlattice potential, $U(\bm{r})$, represents a triangular anti-dot array,

\begin{align}\label{equ:latticePotential}
    U(\bm{r})  &= 2W \sum_{i = 1}^{3} \cos( \bm{g}_{i} \cdot \bm{r}) \\
    \bm{g}_{1} &= \frac{4 \pi}{\sqrt{3} a} ( 0, 1) \nonumber \\
    \bm{g}_{2} &= \frac{4 \pi}{\sqrt{3} a} ( 1/2, \sqrt{3}/2) \nonumber
\end{align}

Where $\bm{g}_{1,2}$ are the basic vectors of the reciprocal lattice and $\bm{g}_{3} = \bm{g}_{2} - \bm{g}_{1}$. The Brillouin zone is shown in Fig. \ref{fig:BZ}. For a triangular lattice with lattice spacing, $a$, the reciprocal vectors have length $|\bm{g}_{i}| = 4 \pi / \sqrt{3} a$. In Eqn. \ref{equ:latticePotential}, the parameter $W$ controls the strength of the potential. To have an anti-dot array we need this to be positive: $W > 0$ (for $W < 0$ the same potential represents an array of dots). Note that a square-lattice potential is symmetric with respect to the replacement $W \rightarrow -W$. The sinusoidal approximation used in Eqn. \ref{equ:latticePotential} is justified by the fact that $U(\bm{r})$ is generated via electrostatic gating. This means that higher harmonics in the potential, $\bm{G} = n \bm{g}_{1} + m \bm{g}_{2}$ for $|n|, |m| > 1$, are suppressed by a factor $e^{- z |\bm{G}|}$, where $z$ is the distance to the gate. For realistic devices, $z$ is sufficiently large and the higher harmonics are negligible.

\begin{figure*}[t!]
    \centering
    \subfloat
    [\label{fig:hoppingt}]
    {\includegraphics[width=0.32\textwidth]{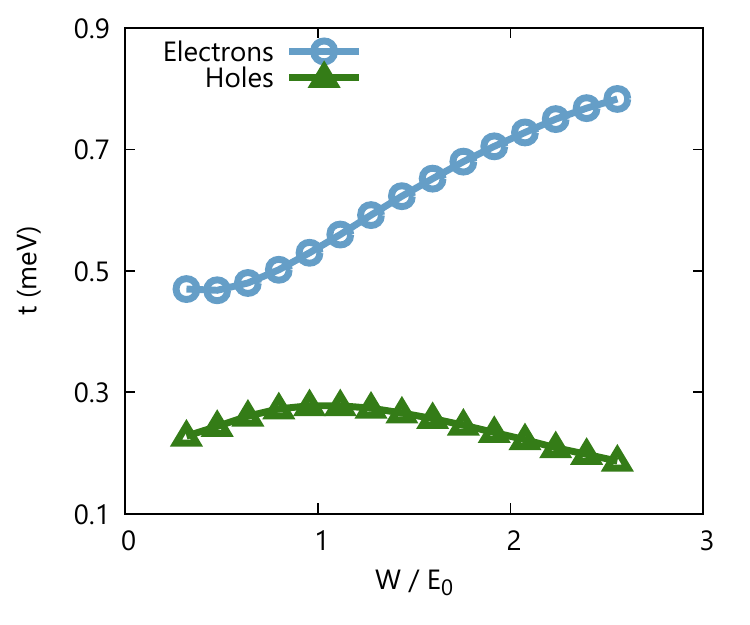}}
\subfloat
    [\label{fig:hubbardU}]
    {\includegraphics[width=0.32\textwidth]{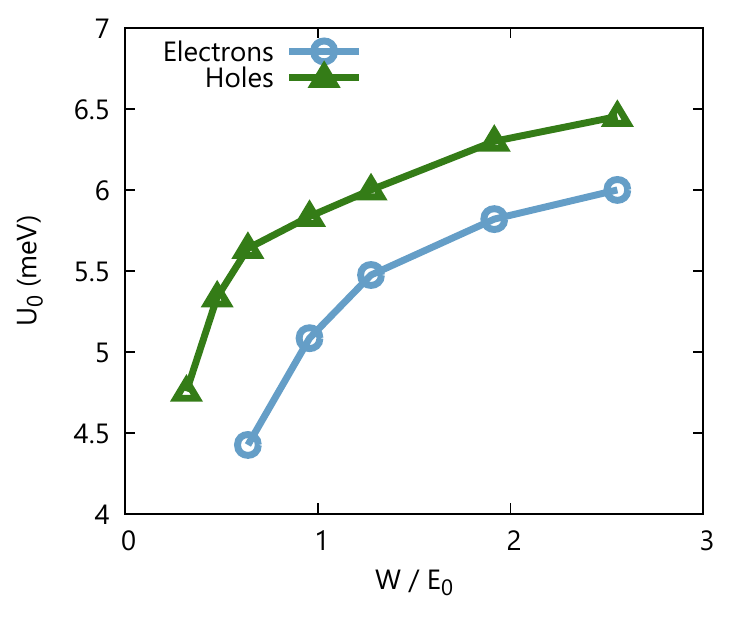}}
\subfloat
    [\label{fig:Uont}]
    {\includegraphics[width=0.32\textwidth]{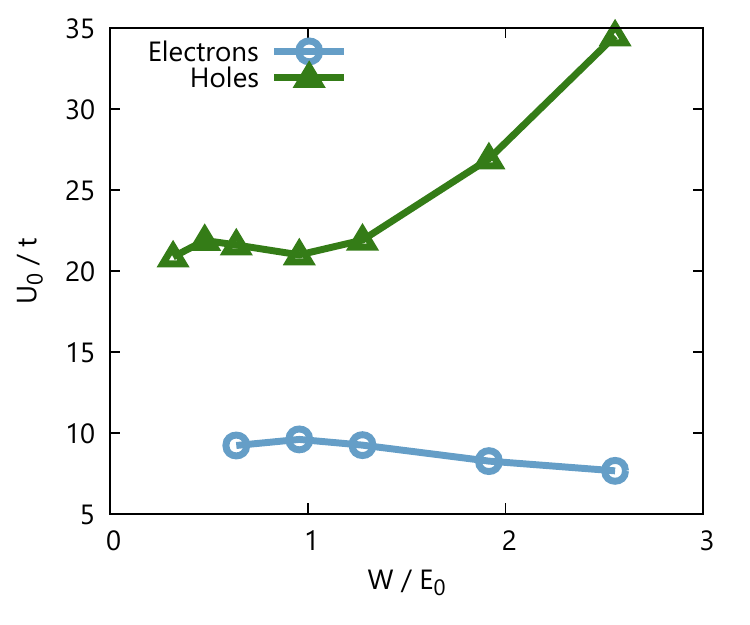}}

\caption{(a) Effective hopping parameter, $t$, as a function of the potential strength $W$. We measure $t$ by equating the total band width of the kagome-like bands (e.g. in figure \ref{fig:AGbandstructure_W=2p5}) to $6t$, the total band width of figure \ref{fig:kagomebandstructure}. Blue corresponds to electrons and green corresponds to holes. (b) Effective Hubbard parameter, $U_{0}$, within the kagome model, as a function of electrostatic potential strength, $W$. (c) The ratio, $U_{0} / t$, between the effective on-site Hubbard repulsion and the effective hopping parameter as a function of $W$.}
\label{fig:hubbard}
\end{figure*}

Numerical diagonalisation of $H$ gives both the band structure and the electron wavefunctions (similar to what was done in Ref. \cite{tkachenkoEffectsCoulombScreening2015}). At a given quasimomentum $\bm{k}$ within the BZ we account for admixture of states with $\bm{k}' - \bm{k} = m \bm{g}_{1} + n \bm{g}_{2}$, where $m$, $n$, are integer numbers. We truncate the Hamiltonian matrix at sufficiently large values of these numbers. The single-particle energy can be expressed in units of $E_{0} = \bm{K}^2 / 2m^{*}$ where $m^{*}$ is the effective mass of GaAs and $ |\bm{K}| = 4 \pi / 3 a$ is the momentum at the $K$-point (see Fig. \ref{fig:BZ} ). This scale determines whether a given value of $W$ is sufficient to strongly reshape the free particle dispersion. For concreteness, we consider a lattice with $a = 80$ nm in a GaAS quantum well ($E_{0} = 1.57$ meV).

At sufficiently large $W$, the dispersion mimics that of the kagome lattice. The 6 lowest energy bands, calculated with $W = 1.0 E_{0}$ and $W = 2.5 E_{0}$, are shown in Fig. \ref{fig:AGplusKagome_bandstructure} (panels \ref{fig:AGbandstructure_W=1p0} and \ref{fig:AGbandstructure_W=2p5} respectively). In the same figure, panel \ref{fig:kagomebandstructure}, we plot the dispersion of a tight-binding model on the kagome lattice with nearest-neighbour, positive hopping parameter $t$, whose Hamiltonian is,

\begin{align*}
    %H &= H_0 + H_{SO} + H_U,\\
    H &= t \sum_{\langle i,j \rangle} c_{i}^{\dag} c_{j}
    %H_{SO}&= \frac{i \Delta_{SO}}{3} \sum_{\langle ij \rangle \alpha\beta} \left(\bm d^1_{ij} \times \bm d^2_{ij}\right)\cdot \bm s_{\alpha\beta} c^\dag_{i\alpha}c_{j\beta},\\
    %H_U &= U_{0} \sum_{i} n_{i, \uparrow} n_{i, \downarrow}.
\end{align*}

Here $c_{i}^{\dag}$ is the creation operator for an electron on site $i$ of the kagome lattice. Comparing figures \ref{fig:AGbandstructure_W=2p5}  and \ref{fig:kagomebandstructure} we conclude that bands 3, 4 and 5 of our TAL reproduce the kagome dispersion very well. Of course, the third band is not perfectly flat (see figure \ref{fig:AG_FB}) but it is close to flat, and, as discussed below, it is even less dispersive for holes. Comparing the total bandwidths of the kagome and TAL dispersions, we can find the effective hopping matrix element $t$, which is plotted in Fig. \ref{fig:hoppingt} as a function of the modulation amplitude, $W$. The typical value of $t$ for electrons is about 0.7 meV.

\begin{figure}[b]
    \centering

\subfloat
    [\label{fig:chargedensity_3}]
    {\includegraphics[width=0.33\textwidth]{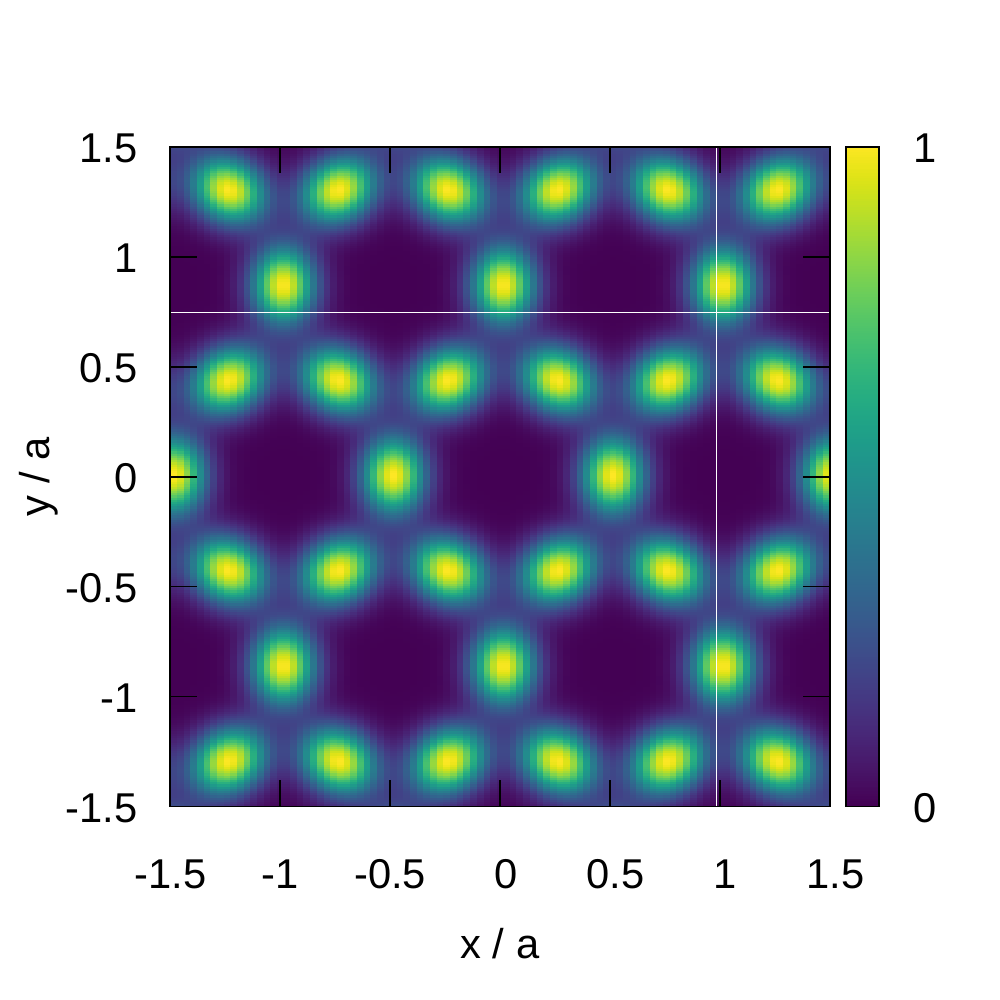}}

\subfloat
    [\label{fig:chargedensity_345}]
    {\includegraphics[width=0.33\textwidth]{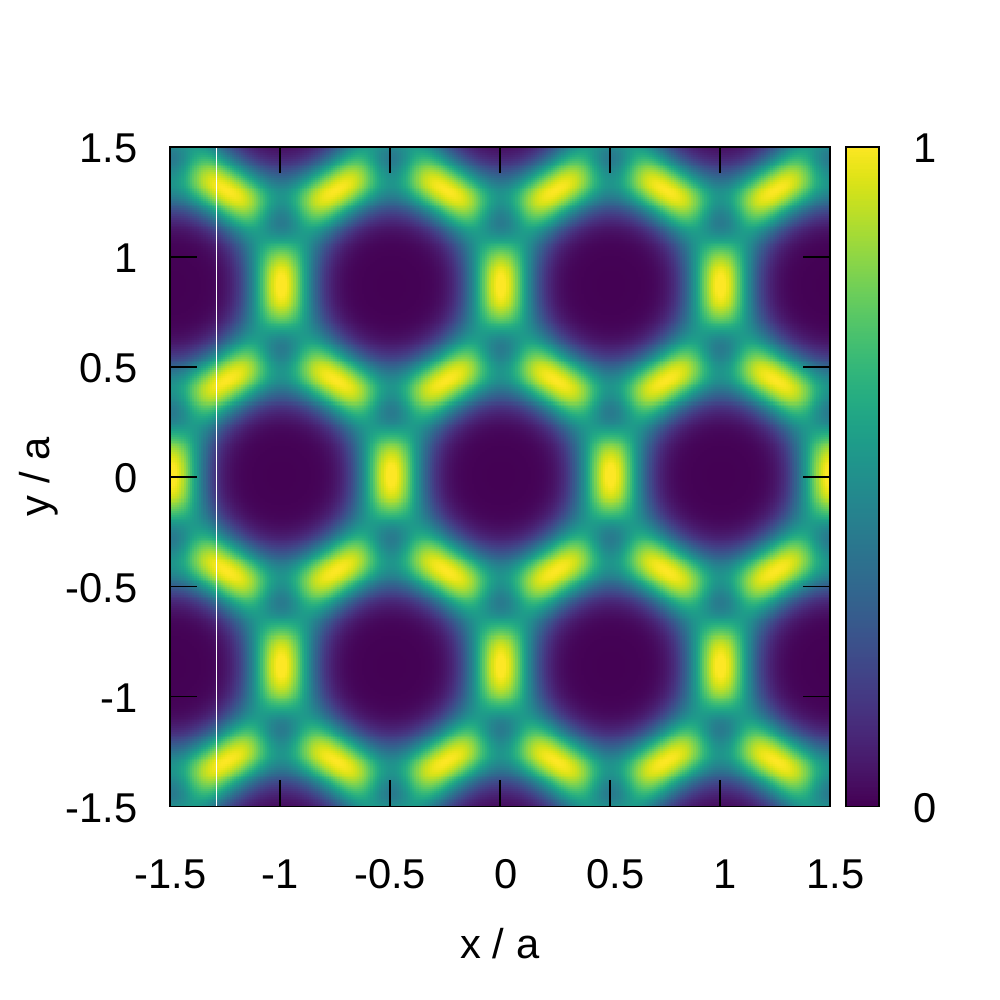}}
\caption{ (a) Charge density of the fully filled flat band in figure \ref{fig:AGbandstructure_W=2p5}. An identical pattern, with reduced intensity, appears for all fractional fillings of the flat band. (b) Total charge density of bands 3, 4, and 5 in figure \ref{fig:AGbandstructure_W=2p5}. In both panels $a = 80$ nm and $W = 2.5 E_{0}$.}
\label{fig:chargedensity}

\end{figure}

Given the Bloch eigenfunctions, $\psi_{m, k}$, of our Hamiltonian, we can calculate the total number density of electrons, $N(\bm{r}) = \sum_{k} |\psi_{m, k} (\bm{r})|^{2}$, for given band, $m$. The map of the number density corresponding to complete filling of the nearly flat band, $m = 3$, is shown in Fig. \ref{fig:chargedensity_3} and the map for complete filling of the kagome-like bands, $m = 3$, 4 and 5, is shown in Fig. \ref{fig:chargedensity_345}. In these maps, the large dark spots (low electron density) are the positions of the anti-dot lattice sites and the bright spots (high electron density) have the symmetry of a kagome lattice. This explains why the triangular anti-dot lattice dispersion emulates that of the kagome model: the kagome-like bands are generated by the formation of an effective kagome lattice within the $m = 3$ energy band.

The bright spots in Fig. \ref{fig:chargedensity_3} are well localized and hence a tight-binding approximation is sensible. Within this approximation, we account for the Coulomb interaction between electrons by calculating the on-site Hubbard repulsion $U_{0}$ for a given kagome lattice site. The calculation of $U_{0}$ proceeds by first treating the electron density profile within a single bright spot as the density $n(\bm{r})$ for a single, localised electron. This requires normalising $n(\bm{r})$ to unity:

\begin{align*}
    \int_{SD} n(\bm{r}) d^{2} r = 1
\end{align*}

Where the region $SD$ is the area around a single bright spot in figure \ref{fig:chargedensity_3}. We can then compute $U_{0}$ as,

\begin{align*}
    U_{0} = \frac{e^{2}}{\varepsilon}
    \int d^{2} \bm{r}_{1} d^{2} \bm{r}_{2}
    \frac{n(\bm{r}_{1}) n(\bm{r}_{2})}{|\bm{r}_{1} - \bm{r}_{2}|}
\end{align*}

Here, $\epsilon$ is the dielectric constant. The plot of $U_{0}$ as a function modulation amplitude $W$ is presented in Fig. \ref{fig:hubbardU}. Our data correspond to GaAs, for which $\epsilon = 12$ (as in all other calculations, $a = 80$ nm). We find a value for $U_{0}$ of around 5-6 meV. In comparison to $t$, we find that $U_{0} / t$ is less than around 10 (see figure \ref{fig:Uont}). Thus, the bands $m = 3$, 4, 5 are mapped to the kagome-lattice Hubbard-model with Hamiltonian:

\begin{align*}
    H &= t \sum_{\langle i,j \rangle} c_i^{\dag}c_j +
    U_{0} \sum_{i} n_{i, \uparrow} n_{i, \downarrow}
\end{align*}

We find that the parameter, $U_{0}$, scales with the lattice constant, $a$, as $U_{0} \sim 1 / \sqrt{a}$ at constant $W$. Since the kinetic energy, $t$, scales as $t \sim 1 / a^{2}$, the ratio $U_{0} / t$ has the following scaling property:

\begin{align*}
    U_{0} / t \sim a^{3 / 2}
\end{align*}

To complete the analysis of the  single electron model we reiterate that the third band of the anti-dot lattice is nearly flat, but, unlike the pure kagome model, it is not perfectly flat. An enlarged plot of the third band of Fig. \ref{fig:AGbandstructure_W=2p5} is given in Fig. \ref{fig:AG_FB}. The minima of this band are at the $K$-points of the Brillouin zone (Fig. \ref{fig:BZ}). Therefore, at low filling of the flat band, the Fermi surface consists of two electron pockets centered at $\bm{K}$ and $\bm{K}'$. The band dispersion near these points is quadratic, $E = \bm{q}^{2} / 2 m^{*}$, and the effective mass, $m^{*}$, (discussed below) is plotted versus $W$ in Fig. \ref{fig:effMass}.

\section{Hole Band Structure, Comparison with Electrons}\label{sec:holes}

The effective kagome model applies to both holes and electrons, however, the hole superlattice has some significant advantages when it comes to experimental realisations. In this section we demonstrate, using band structure calculations, that holes in a TAL require a much weaker modulation to achieve the same flatband width and interaction strength as electrons.

\subsection{Technique for computing hole band structure}

In contrast to electrons, holes have strong spin-orbit coupling. Energy bands for the hole system can thus be derived from the Luttinger Hamiltonian \cite{luttingerQuantumTheoryCyclotron1956} $H_{L}$, which accounts for spin-orbit coupling. The total Hamiltonian is,

\begin{align}\label{equ:holeHamiltonian}
    H = H_{L} + w(z) + U(x,y)
\end{align}

Where $w(z)$ is a confining potential oriented perpendicular to the plane of the artificial lattice (again, defined by $U(x,y)$ from Eqn. \ref{equ:latticePotential}). Specifically, $w(z)$ is taken to be an infinite square well of width $d = 15$ nm. The Luttinger Hamiltonian can be written as (Eqn. 6 in Ref.\cite{miserevDimensionalReductionLuttinger2017}),

\begin{align}\label{equ:luttingerHamiltonian}
    H_{L} =& H_{0} + V \\
    H_{0} =&
    \left( \gamma_{1} + \frac{5}{2} \bar{\gamma} - 2 \bar{\gamma} S_{z}^{2} \right)
    \frac{p_{\perp}^{2}}{2 m_{e}} + \nonumber \\
    &\left( \gamma_{1} - \frac{5}{4} \bar{\gamma} +   \bar{\gamma} S_{z}^{2} \right)
    \frac{\bm{p}_{\parallel}^{2}}{2 m_{e}} \nonumber \\
    V =&
    \frac{- \bar{\gamma}}{4 m_{e}}
    [ p_{+}^{2} S_{-}^{2} + p_{-}^{2} S_{+}^{2} + \nonumber \\ &
    2 p_{z} p_{+} \{ S_{z}, S_{-} \} + 2 p_{z} p_{-} \{ S_{z}, S_{+} \}]
    \nonumber
\end{align}

Where $p_{\pm} \equiv p_{x} \pm i p_{y}$ and $S_{i}$ are the $4 \times 4$ spin matrices for a spin $3/2$ particle. For simplicity we used the spherical approximation, $\gamma_{2} \approx \gamma_{3}$, with $\bar{\gamma} \equiv (2\gamma_{2} + 3\gamma_{3})/5$. The parameters $\gamma_{1}$ and $\bar{\gamma}$ are taken to be $6.85$ and $2.58$ respectively, values corresponding to GaAs. Our approach to solving this problem begins with defining an orthonormal set of basis wavefunctions $\Psi_{k, n, \sigma}$:

\begin{align}\label{equ:basisVector}
    \Psi_{k, n, \sigma}(\bm{r}) =
    \phi_{n}(z) \psi_{k}(x,y) \chi_{\sigma}
\end{align}

Where, $\phi_{n}(z)$ is an eigenfunction of the square well, $\psi_{k}(x,y)$ is a plane wave (eigenfunction of the momentum operator) and $\chi_{\sigma}$ is a spin $3/2$ spinor (eigenfunction of $S_{z}$). As an example, for $\sigma = + 3 / 2$ and arbitrary $n = \text{odd}$ and $k$,

\begin{align*}
    \Psi_{k, n, \sigma=3/2}(\bm{r}) =
    \sqrt{\frac{2}{d}}
    \cos ( n \pi z / d )
    \frac{1}{\sqrt{A}}
    e^{ i \bm{k} \cdot \bm{r}_{\parallel} }
    \begin{bmatrix}
        1 \\ 0 \\ 0 \\ 0
    \end{bmatrix}
\end{align*}
Here, $A$ is the area of the sample and when $n$ is even $\cos$ changes to $\sin$. The wavefunction is also zero for $|z| > d/2$ (i.e. in the region forbidden by $w(z)$). We can now compute the matrix elements of the Hamiltonian (Eqn. \ref{equ:holeHamiltonian}) in the basis defined by these functions,

\begin{align*}
    H_{ \genfrac..{0pt}{2}{ k_{i}, n, \sigma}{ k_{j}, m, \tau} } =
    \bra{ \bm{k}_{i}, n, \sigma} H
    \ket{ \bm{k}_{j}, m, \tau}
\end{align*}

With regards to $x$ and $y$ dependence, the non-modulated part of the Hamiltonian (in Eqn. \ref{equ:holeHamiltonian}), $H_{L} + w(z)$, contains only $p_{x}$, $p_{y}$ and has no explicit dependence on the variables $x$, $y$. It follows that this part of $H$ is diagonal in the index $k_{i}$ since the basis vector, $\Psi$, is a plane-wave in the $x$-$y$ plane with wave vector $\bm{k}_{i}$. As for the modulated part of the Hamiltonian, it is well known that a periodic potential, $U(x,y)$, has non-zero matrix elements only between plane waves which differ in momentum by a reciprocal lattice vector. We can thus define a quasi-momentum $\bm{k}$ and let $\bm{k}_{i} = \bm{k} + \bm{G}_{i}$ where $\bm{G}_{i}$ is a reciprocal lattice vector. The matrix elements of the Hamiltonian can then be written as,

\begin{align}\label{equ:holeMatrixElements}
    H_{ \genfrac..{0pt}{2}{ i, n, \sigma}{ j, m, \tau} }(\bm{k}) =
    \bra{ \bm{k} + \bm{G}_{i}, n, \sigma} H
    \ket{ \bm{k} + \bm{G}_{j}, m, \tau}
\end{align}

We are now in a position to compute these matrix elements explicitly. The expression for $H_{0}$ in Eqn. \ref{equ:luttingerHamiltonian} contains only $S_{z}$, $p_{z}^{2}$ and $p_{x}$, $p_{y}$. Our basis vectors (Eqn. \ref{equ:basisVector}) are, by design, eigenvectors of $H_{0} + w(z)$. This part of the Hamiltonian is thus diagonal in all indices $i$, $n$ and $\sigma$; its matrix elements are given by $H_{0}$ in Eqn. \ref{equ:luttingerHamiltonian} with $\bm{p}_{\parallel} = \bm{k} + \bm{G}_{i}$, $p_{z}^{2} = (n \pi / d)^{2}$ and $S_{z} = \sigma$. Thus,

\begin{align*}
    ( H_{0} + w(z) )_{ \genfrac..{0pt}{2}{ i, n, \sigma}{ j, m, \tau} } =& \
    \delta_{i,j} \delta_{n, m} \delta_{\sigma, \tau} \times \\
    &
    \left[
    \left( \gamma_{1} + \frac{5}{2} \bar{\gamma} - 2 \bar{\gamma} \sigma^{2} \right) \right.
    \frac{( n \pi / d )^{2}}{2 m_{e}} + \nonumber \\
    & \left. \left( \gamma_{1} - \frac{5}{4} \bar{\gamma} +   \bar{\gamma} \sigma^{2} \right)
    \frac{( \bm{k} + \bm{G}_{i} )^{2}}{2 m_{e}} \right] \nonumber \\
\end{align*}

The periodic potential $U(x,y)$ has matrix elements,

\begin{align*}
    ( U )_{ \genfrac..{0pt}{2}{ i, n, \sigma}{ j, m, \tau} } =
    W \sum_{\alpha = 1}^{3} \delta( \bm{G}_{j} - \bm{G}_{i} \pm \bm{g}_{\alpha} ) \delta_{n, m} \delta_{\sigma, \tau}
\end{align*}

Lastly, the operator, $V$, in Eqn. \ref{equ:luttingerHamiltonian} contains $p_{x}$ and $p_{y}$ which are diagonal in the indices $i$, $j$ and will thus be replaced by $\delta_{i, j} ( \bm{k} + \bm{G}_{i} )_{x,y}$. It also contains $S_{x, y, z}$, whose matrix elements in the indices $\sigma$, $\tau$ are given, for example, in the textbook Ref.\cite{landau_Volume_3}. The only remaining term is $p_{z}$, which has matrix elements:

\begin{align*}
    &(p_{z})_{ \genfrac..{0pt}{2}{ i, n, \sigma}{ j, m, \tau} } =
    \delta_{i, j}
    \delta_{\sigma, \tau} \times \\ &
    \begin{cases}
        0, & n = m \\
        - \frac{2 m i}{d}
        \left[
            \frac{\sin((n - m) \pi / 2)}{(n - m)} +
            \frac{\sin((n + m) \pi / 2)}{(n + m)}
        \right],
        & \genfrac..{0pt}{2}{n = \text{odd}}{ m = \text{even}} \\
        + \frac{2 m i}{d}
        \left[
            \frac{\sin((n - m) \pi / 2)}{(n - m)} -
            \frac{\sin((n + m) \pi / 2)}{(n + m)}
        \right],
        & \genfrac..{0pt}{2}{ n = \text{even}}{ m = \text{odd} }
    \end{cases}
\end{align*}

The complete matrix for $V$ is then,

\begin{align*}
    ( V )_{ \genfrac..{0pt}{2}{ i, n, \sigma}{ j, m, \tau} } = & \
    \frac{- \bar{\gamma}}{4 m_{e}}
    \delta_{i, j} \times \\
    [ &(k + G_{i})_{+}^{2} (S_{-}^{2})_{\sigma, \tau} + \\ &
       (k + G_{i})_{-}^{2} (S_{+}^{2})_{\sigma, \tau} + \\ &
    2 ( p_{z} )_{n, m} ( k + G_{i} )_{+}
        \{ S_{z}, S_{-} \}_{\sigma, \tau} + \\ &
    2 ( p_{z} )_{n, m} ( k + G_{i} )_{-}
        \{ S_{z}, S_{+} \}_{\sigma, \tau}]
    \nonumber
\end{align*}

\begin{figure}[t]
\centering
\includegraphics[width=0.30\textwidth]{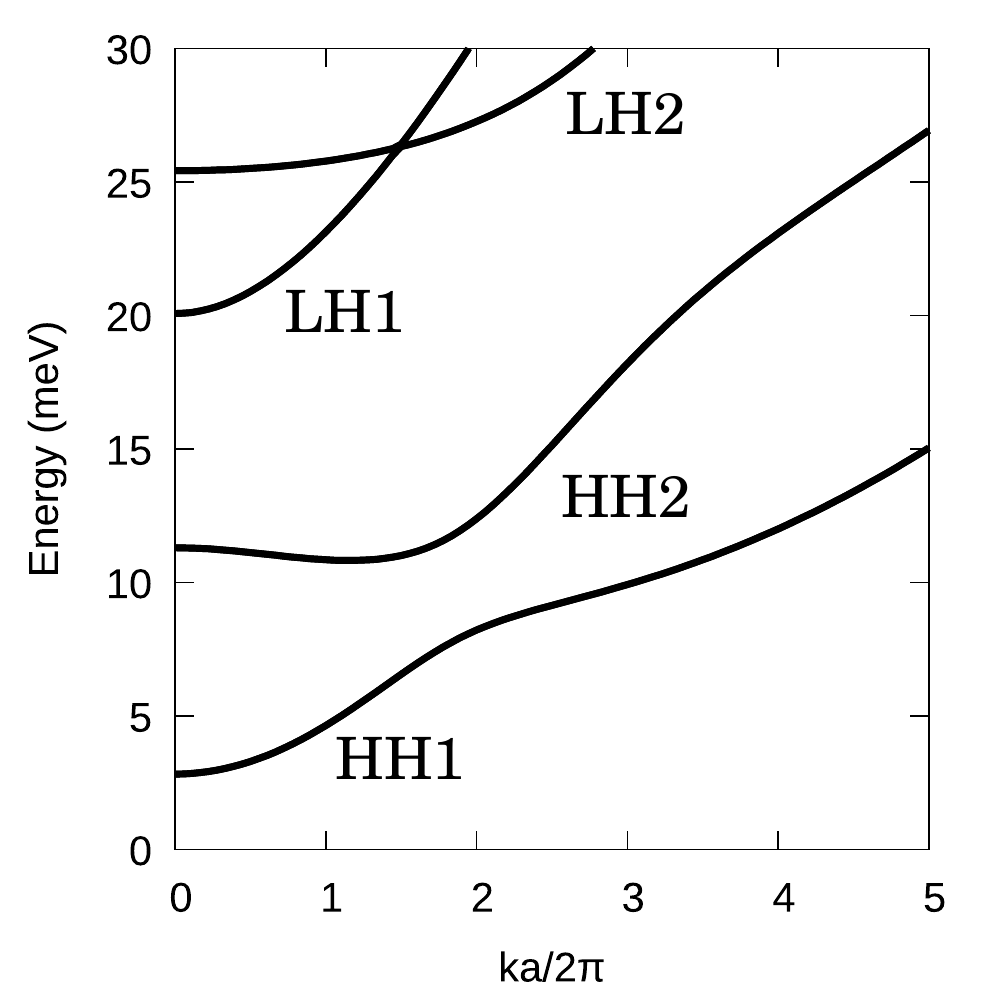}
\caption{Bare dispersion for 2D holes in GaAs with an infinite square confining potential. Here, $d = 15$ nm and we measure momentum in units of $2 \pi / a$ for $a = 80$ nm (the standard lattice constant throughout this work). This dispersion is isotropic.}
\label{fig:hole_bare}
\end{figure}

We now have an explicit expression for all the matrix elements of $H = H_{0} + w(z) + V + U(x,y)$ in the basis defined by Eqn. \ref{equ:basisVector}. Energy levels and eigenvectors of $H$ can then be determined by numerical diagonalisation, provided we truncate the basis. The truncation procedure amounts to choosing a maximum value for $n$ and a finite set of reciprocal lattice vectors, $\bm{G}_{i}$, to include in the basis. To determine appropriate values, we increased the size of the basis until all energy levels of interest converged. The energy levels in the absence of any periodic modulation (shown in Fig. \ref{fig:hole_bare}), computed using this method, agree with previous calculations\cite{miserevDimensionalReductionLuttinger2017, winklerSpinOrbitCoupling2003}.

\begin{figure*}[t]
    \centering
\subfloat
    [\label{fig:holeBands_W=0p25} $W = 0.25 E_{0}$]
    {\includegraphics[width=0.25\textwidth]{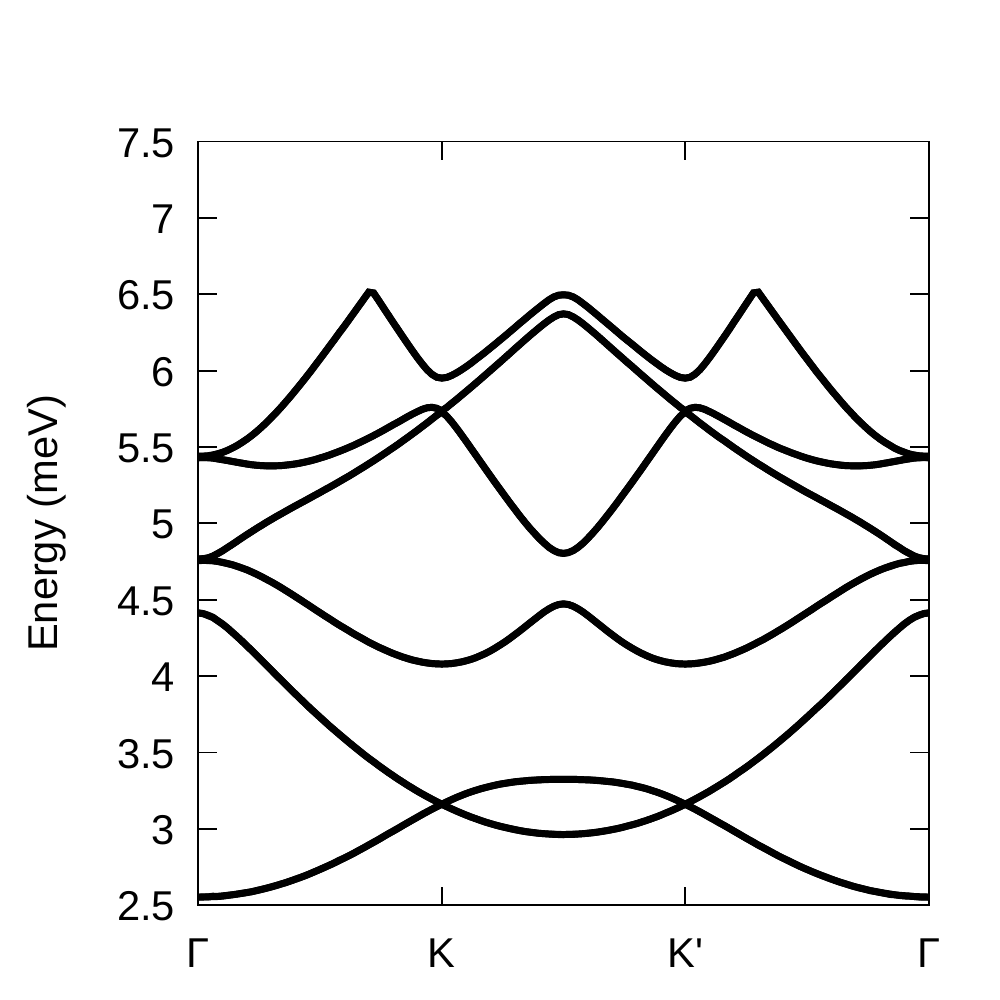}}
\subfloat
    [\label{fig:holeBands_W=0p50} $W = 0.50 E_{0}$]
    {\includegraphics[width=0.25\textwidth]{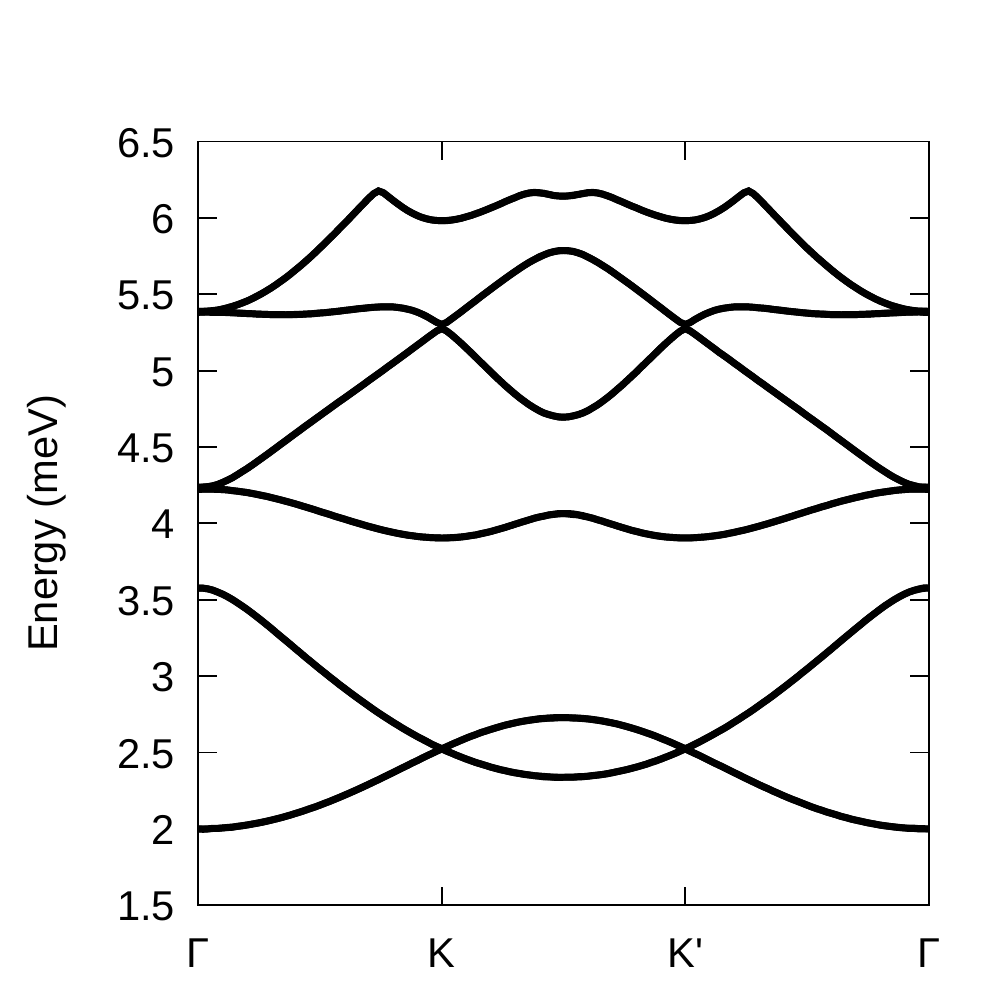}}
\subfloat
    [\label{fig:holeBands_W=1p00} $W = 1.00 E_{0}$]
    {\includegraphics[width=0.25\textwidth]{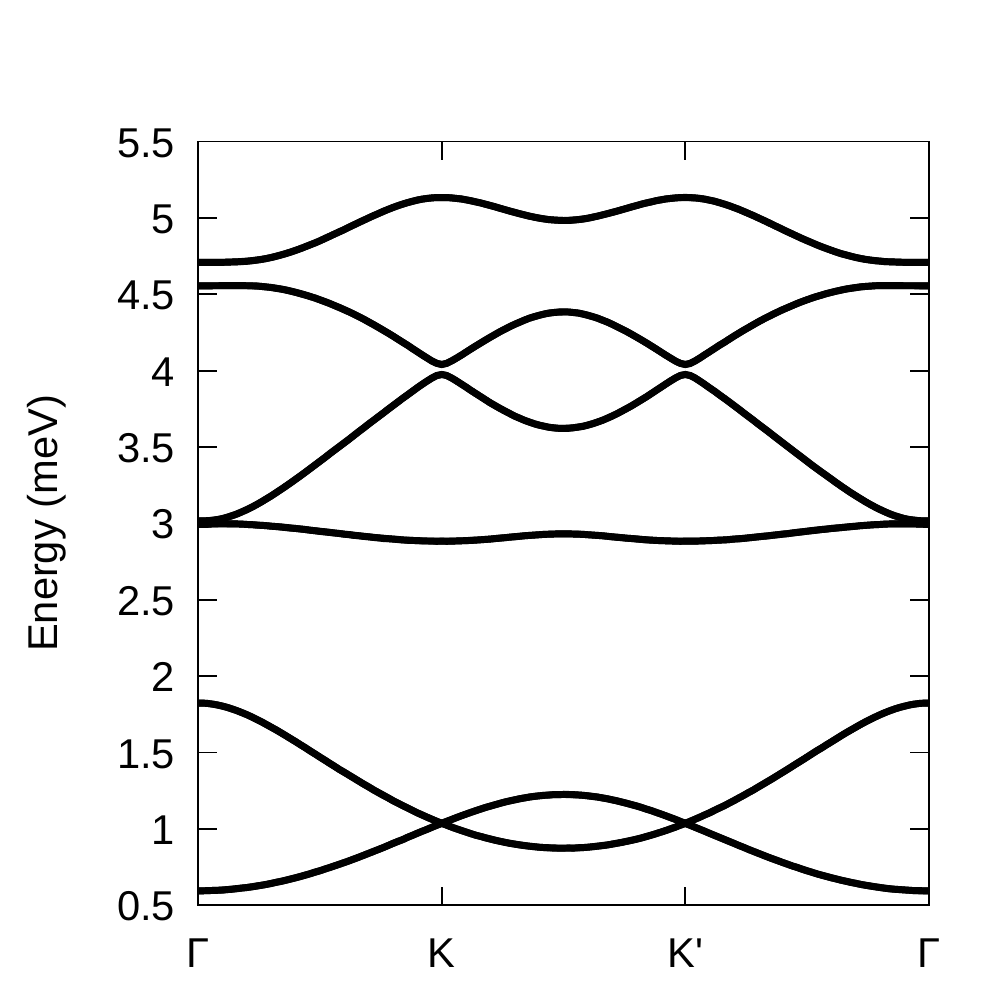}}
\subfloat
    [\label{fig:holeBands_W=2p50} $W = 2.50 E_{0}$]
    {\includegraphics[width=0.25\textwidth]{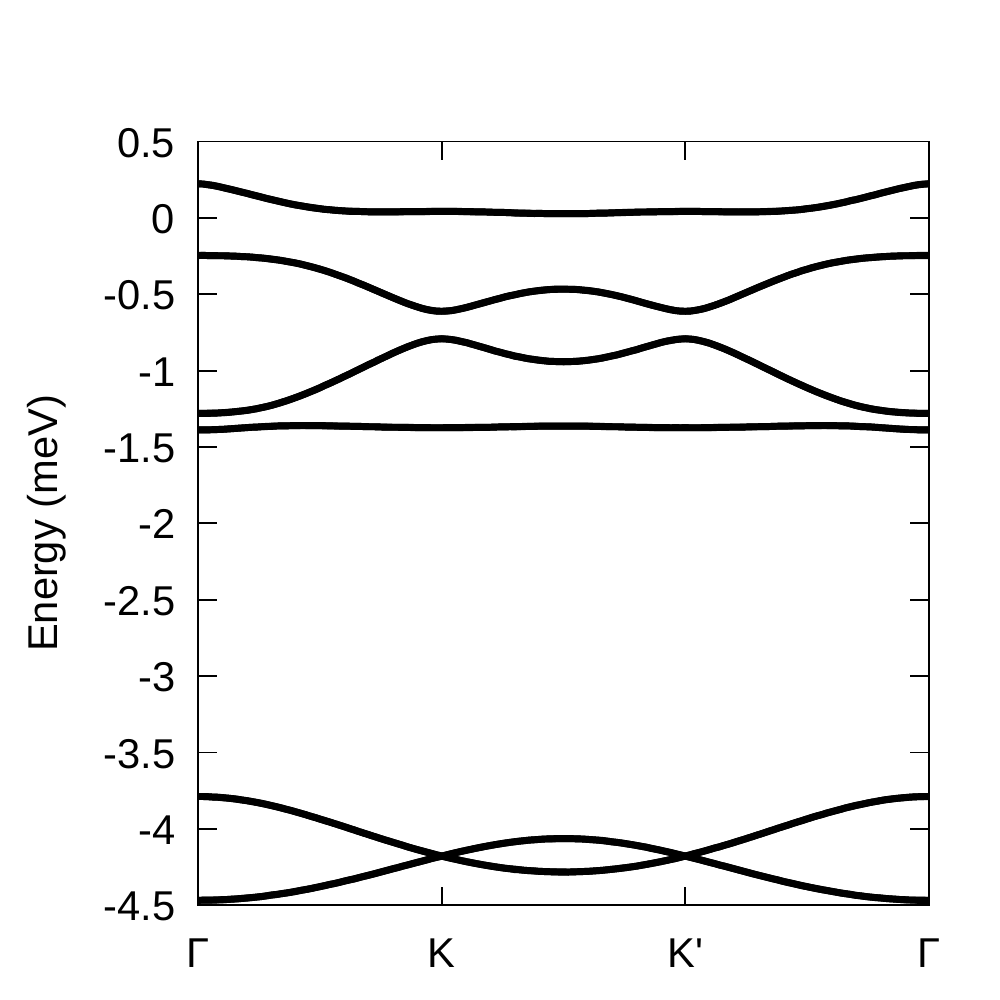}}

\caption{Energy bands for holes, derived from the Hamiltonian in Eqn. \ref{equ:holeHamiltonian}, for various values of $W$ at $a = 80$ nm and quantum well width $d = 15$ nm. Each hole band is doubly degenerate.}
    \label{fig:holeBands}
\end{figure*}

\subsection{Hole band structure results}

\begin{figure}[t]
\centering

\subfloat
    [\label{fig:holeBS_d=20}]
    {\includegraphics[width=0.240\textwidth]{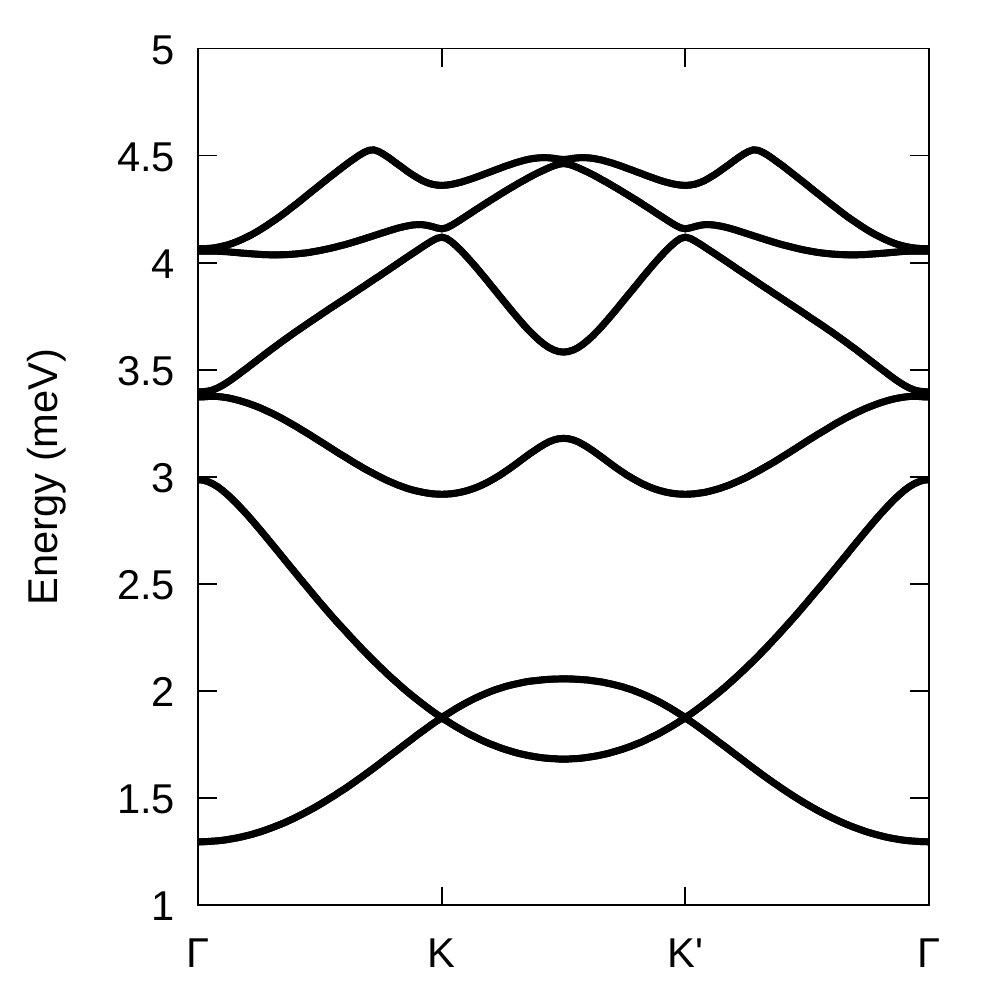}}
\subfloat
    [\label{fig:holeBS_d=40}]
    {\includegraphics[width=0.240\textwidth]{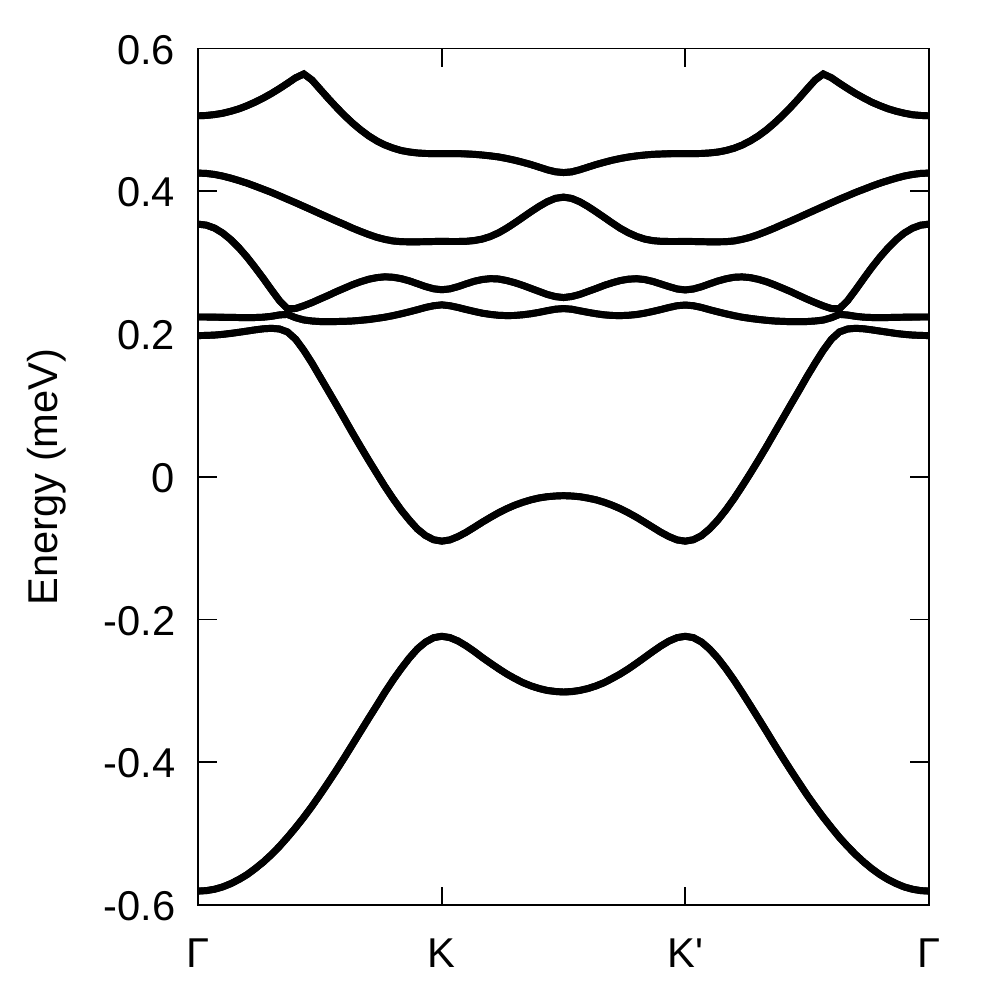}}

\caption{Lowest 6 bands of the hole band structure for $d = 20$ nm (left) and $d = 40$ nm (right). In both panels $W = 0.5 E_{0}$ and $a = 80$ nm. In panel (b) the gap between the two lowest bands is $14\%$ of the total width of those bands.}
\label{fig:hole_large_d}
\end{figure}

Examples of the hole mini bands are presented in figure \ref{fig:holeBands}. In these calculations the strength of the spin-orbit interaction is governed by the ratio $d / a$ between the well width, $d$, and the lattice constant, $a$. We have considered the weak to moderate spin-orbit regime in this work, with $d / a \approx 0.2$. The hole dispersion has a similar form to that of an equivalent electron system, the two lowest energy bands are Dirac-like and the next three bands are kagome-like (see figure \ref{fig:holeBands_W=1p00}, for example). There are some key differences, however. Holes in an unmodulated device have a non-parabolic dispersion (Fig. \ref{fig:hole_bare}) and, in general, will have a different effective mass than electrons. Because of this, the effective mass for holes is a function of momentum and does not have a single, well defined value. Roughly speaking, however, the unmodulated effective mass for holes is three times larger than that for electrons. The second key difference is the spin-orbit interaction. As mentioned above, holes in GaAs have a non-negligible spin-orbit interaction. The presence of this interaction introduces gaps at the Dirac points and at the flat band that do not exist in the electron band structure. Note that for larger values of $d / a$ this causes a much more significant reshaping of the energy bands (Fig. \ref{fig:hole_large_d}). For example, at $d / a = 0.5$ and $W = 0.5 E_{0}$ the kagome-like bands disappear and the graphene-like bands remain with a large gap relative to the total band width (Fig. \ref{fig:holeBS_d=40}). This regime is interesting in connection with artificial topological insulators \cite{SushkovNeto2013, ScammellSushkov2019}, but is not the main focus of this work.

\begin{figure*}[ht]
\centering
\subfloat[\label{fig:effMass}]
         {\includegraphics[width=0.32\textwidth]{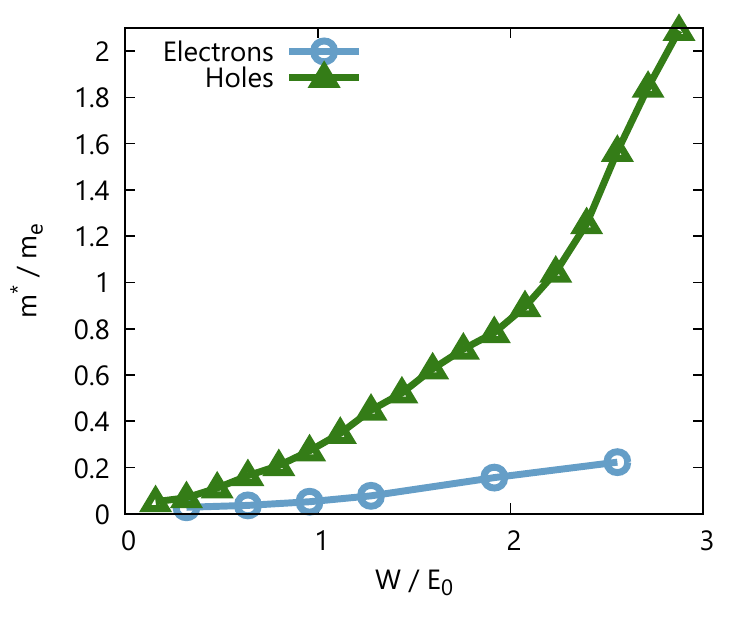}}
\subfloat[\label{fig:stoner}]
         {\includegraphics[width=0.32\textwidth]{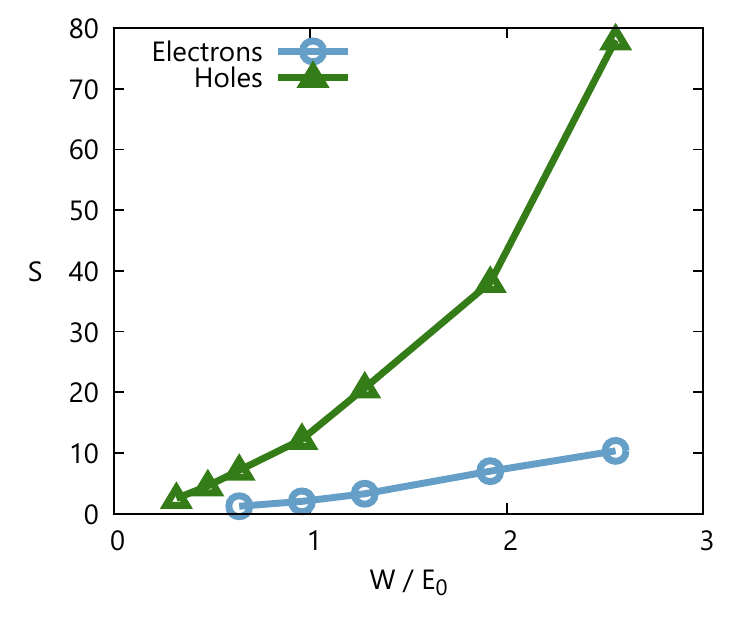}}

\caption{(a) Effective mass of the flat band (defined around the $K$-points in Fig. \ref{fig:AG_FB}) as a function of potential strength $W$ for electrons and holes. The quantity $1/m^{*}$ is roughly proportional to the width of the flatband. (b) Plot of the Stoner parameter, $S$ (Eqn. \ref{equ:stoner}), as a function of $W$ for both electrons and holes.}
\label{fig:effMassStoner}
\end{figure*}

Our central finding with regards to the hole band structure is that the kagome-like bands, including the flat band, develop at a much weaker modulation than for electrons. For example, compare figure \ref{fig:AGbandstructure_W=1p0} for electrons at $W = E_{0}$ and figure \ref{fig:holeBands_W=1p00} for holes at the same $W$. The hole bands have a well formed kagome-like dispersion while the electron bands do not. There is also a factor 7 difference in the band width of the third band between holes and electrons, with holes having the much flatter band. The degree of band flatness is captured by the curvature around the minimum of the flat band (see Fig. \ref{fig:AG_FB}). Since this part of the dispersion is parabolic we can describe it by an effective mass, $\varepsilon(k) = k^{2} / 2 m^{*}$. We compare this effective mass for holes and electrons in figure \ref{fig:effMass}, from which it can be seen that $m^{*}$ is 4 to 10 times larger in the hole flat band than in the electron flat band. Figure \ref{fig:effMass} also shows that the effective mass $m^{*} = 0.2 m_{e}$, for example, is reached at $W \approx 2.5 E_{0}$ for electrons and $W \approx 0.6 E_{0}$ for holes. The third hole band is not just flatter in absolute units, it is also flatter relative to the total width of the kagome-like bands (proportional to $t$, Fig. \ref{fig:hoppingt}). From figure \ref{fig:hoppingt} it can be seen that $t$ is 2 to 9 times smaller for holes. This decrease in total bandwidth is compensated by a greater decrease in the width of the flat band.

To conclude this section we note that one of the experimental challenges in producing artificial superlattices is generating a strong periodic modulation, sufficient to significantly restructure the energy bands of the 2D system. Our calculations show that hole systems require a much smaller modulation strength to induce flat bands, and hence access strongly correlated phases, than equivalent electron systems. This is the central conclusion of our hole band structure calculation.

\section{Possible Strongly Correlated Phases Within the Effective Kagome Model}\label{sec:strong}

In the present section we discuss some of the possible strongly correlated phases which could arise in TALs. Here we use a strong Coulomb coupling expansion, i.e. the emergent correlated phases are those that minimise the Coulomb energy.

\subsection{Commensurate Charge Density Waves}

The existence of an underlying kagome lattice and the large value of $U_{0} / t$ (Fig. \ref{fig:Uont}) imply that the on-site localization of electrons (or holes) is highly likely. This localization eliminates on-site Coulomb repulsion, however, longer range  Coulomb repulsion is still present. For example, the distance between nearest kagome sites is $a/2=40$nm. The nearest-site Coulomb repulsion is then very significant,

\begin{eqnarray}
V =\frac{e^2}{\epsilon a/2} \approx 2.7 \ \text{meV} \ . \nonumber
\end{eqnarray}

This longer range Coulomb repulsion can lead to ordering of the localized electrons (or holes). Thus, in this subsection, we consider the corresponding possible commensurate CDWs.

These CDWs would exist on the kagome lattice at certain filling fractions of the kagome-like bands in figure \ref{fig:AGbandstructure_W=2p5} (or \ref{fig:holeBands_W=1p00}), each filling fraction having a set of possible CDW patterns. Here we consider, for the purposes of illustration, the filling fractions $n = 1 n_{0}, (3/2) n_{0}, (4/3) n_{0}, \ \text{and} \ 2 n_{0}$ measured relative to complete filling of the lowest two bands and in units of $n_{0} = 1 / A_{cell}$. Thus $n = 1n_0$ corresponds to half filling of the flat band and $n = 2n_0$ corresponds to full filling of the flat band.

\begin{figure*}[pt]
%\centering
\includegraphics[width=0.8\textwidth]{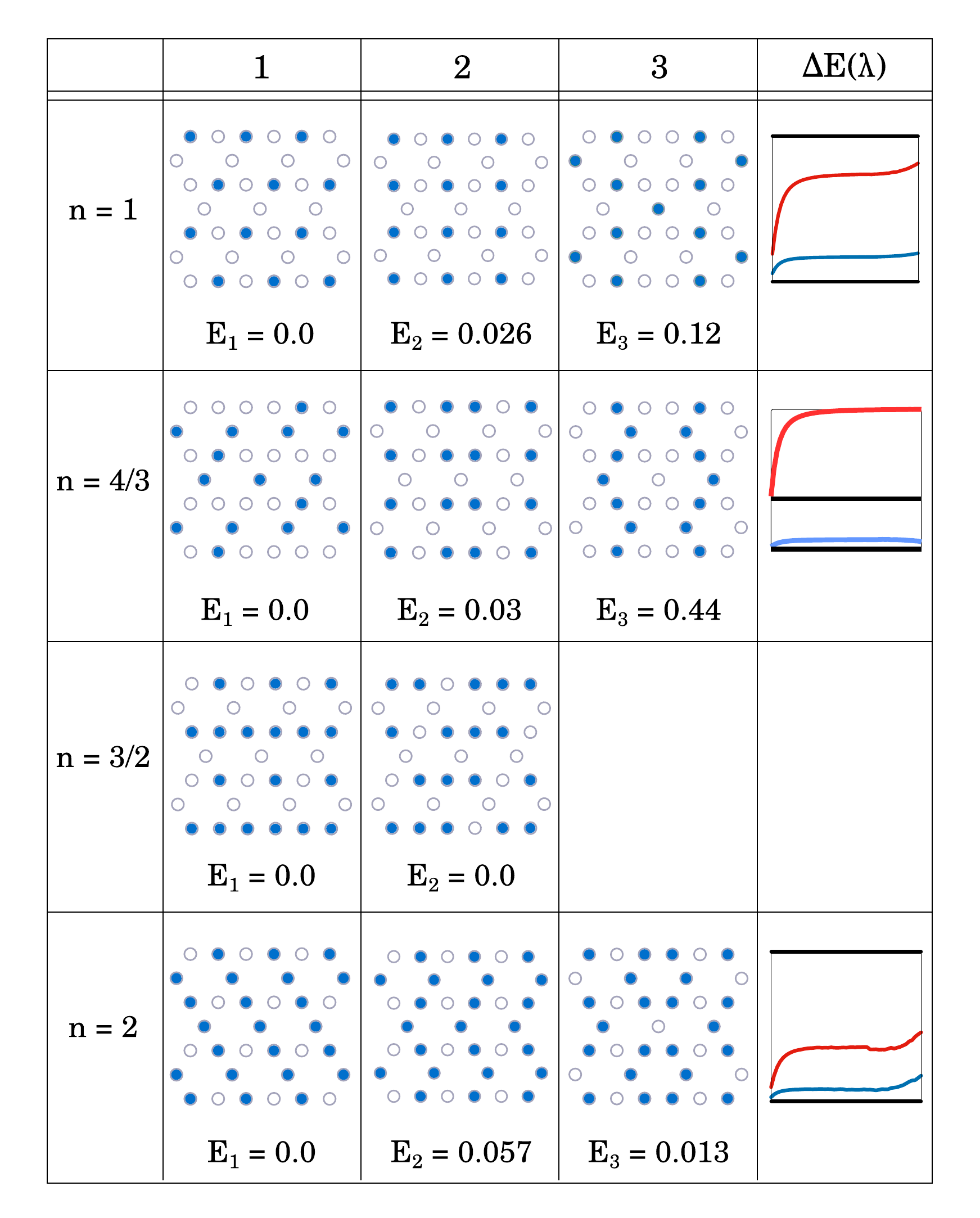}
\caption{Summary of results for the average energy per electron of each CDW pattern. Electron density, $n$, is given in units of one electron per unit cell. Each energy, $E_{i}$, was computed for the corresponding pattern pictured in each cell and the patterns are numbered in order of increasing energy. Energy is given in meV measured relative to the lowest energy configuration. The rightmost column plots the energy of each pattern as a function of $\lambda$ (the screening length) relative to the lowest energy configuration $E_{1}$. The flatband width is shown in black and $\lambda$ varies from $0$ to $5 a$.}
\label{fig:table}
\end{figure*}

We can now catalog the set of CDW phases that are possible within the flat band. For each filling fraction we found periodic patterns of occupied kagome-lattice sites which give the correct amount of charge per unit cell. These are presented in figure \ref{fig:table}. There can, in general, be more than one pattern at each filling fraction. For each of these patterns, the Coulomb energy per electron is,

\begin{align*}
    E = \frac{1}{2 N_{sites}}
    \sum_{ \genfrac{.}{.}{0pt}{1}{i, j = 1}{j \neq i} }^{N_{sites}}
    \frac{ e^{2}}{\epsilon r_{ij}}
    e^{- r_{ij} / \lambda}
\end{align*}

Where $i$ and $j$ represent occupied sites in the CDW, $r_{ij}$ is the distance between sites and $\lambda \approx 3a$ is the screening length (due to screening by image charges in the metallic gate). There is also $N_{sites} \rightarrow \infty$, which is the number of occupied lattice sites. To avoid the infinite summation, we define a smaller set of $M$ occupied sites. This block has to be defined such that it can be repeated, periodically, to reconstruct the full CDW pattern. The energy we compute is,

\begin{align*}
    E = \frac{1}{2M}
    \sum_{ \genfrac{.}{.}{0pt}{1}{i = 1, M}{j \neq i} }
    \frac{e^{2}}{\epsilon r_{ij}}
    e^{- r_{ij} / \lambda}
\end{align*}

The right column of figure \ref{fig:table} shows this energy as a function of $\lambda$ measured relative to the lowest energy configuration. We find that different CDW patterns at the same density can be distinguished by energies which differ on the order of the electron flatband width and the first column of figure \ref{fig:table} identifies the pattern with minimal energy. Since the CDW phase is insulating, its signature in transport measurements will be a maximal value of $R_{xx}$ when particle density is tuned to one of the values given in figure \ref{fig:table}. If we measure the particle density from the bottom of the lowest energy band then these values are $n_{tot} = 9 \times 10^{10} \text{cm}^{-2}, \ 9.6 \times 10^{10} \text{cm}^{-2}, \ 9.9 \times 10^{10} \text{cm}^{-2}$ and $10.8 \times 10^{10} \text{cm}^{-2}$.

\subsection{Mott insulator}

The Mott insulating phase is related to the CDW phase but occurs exactly at half filling of the kagome lattice (i.e. one particle on each kagome lattice site, $n = 3n_0$). This does not occur within the flat band but within the band directly above it, i.e. at half filling of the second kagome-like band (Figs. \ref{fig:AGbandstructure_W=2p5} and \ref{fig:holeBands_W=1p00}). The experimental signature for this phase is the same as for charge density waves except the density at which this occurs is $n_{tot} = 12.6 \times 10^{10} \text{cm}^{-2}$. The effective antiferromagnetic exchange between nearest sites in the Mott insulator is $J=4t^2/U_0\sim 0.1$meV.

\section{Possible Weak coupling phases}\label{sec:weak}

In this section we continue to analyse possible quantum phases within the flatband (see Fig.\ref{fig:holeBands_W=0p25}) using the itinerant picture. We thus account for the Coulomb interaction perturbatively.

%In the previous sections, we have argued that the kagome flatband hosts sufficiently strong Coulomb interactions (relative to kinetic energy) for the system to be in a strongly correlated phase. On the other hand, it is worth understanding which phases emerge from the single particle band structure by accounting for the Coulomb interaction perturbatively and we discuss this approach in the present section. For both electrons and holes, 

The weak Coulomb coupling regime is reached by considering a weak potential modulation, characterised by $W / E_{0}$. For example, in the hole gas, $W / E_{0} \lesssim 0.25$ corresponds to weak Coulomb coupling in the kagome flatband. We see from Fig. \ref{fig:holeBands_W=0p25} that for $W / E_{0} = 0.25$, the flatband shows significant dispersion, whereas for $W / E_{0} = 1$, this band is nearly dispersionsless.

\subsection{Ferromagnetism}

We have pointed out above that because of the large value of $U_0/t$ the localization of electrons/holes is likely. Nevertheless it is instructive to consider the itinerant picture as well. This analysis is probably more relevant to the relatively weak superlattice
modulation, $W/E_0 \lesssim 1$.

\begin{figure*}[t]
\centering
\subfloat
    [\label{fig:nesting}]
    {\includegraphics[width=0.275\textwidth]{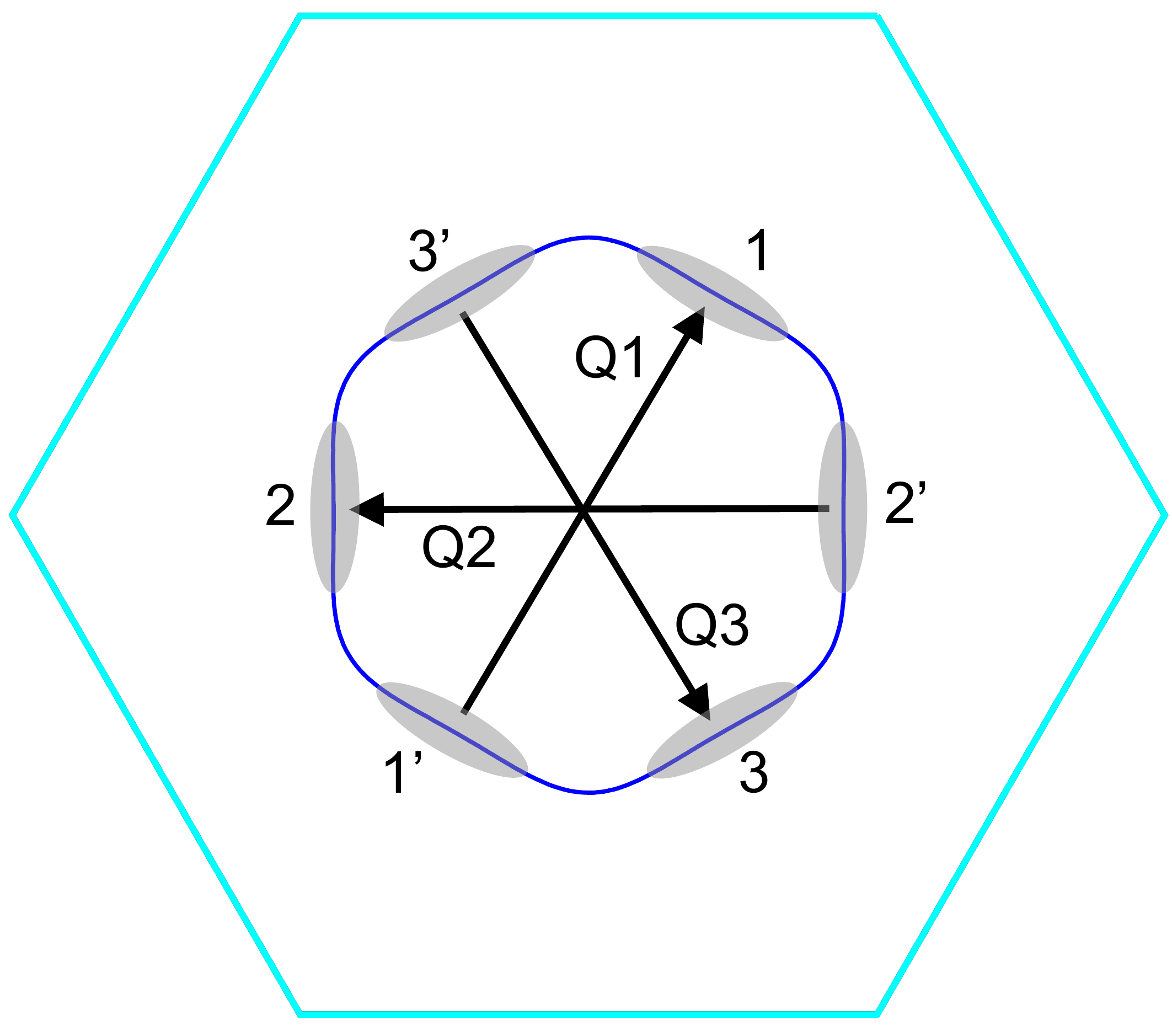}}\hspace{0.25cm}
\subfloat
    [\label{fig:miniBZ}]
    {\includegraphics[width=0.272\textwidth]{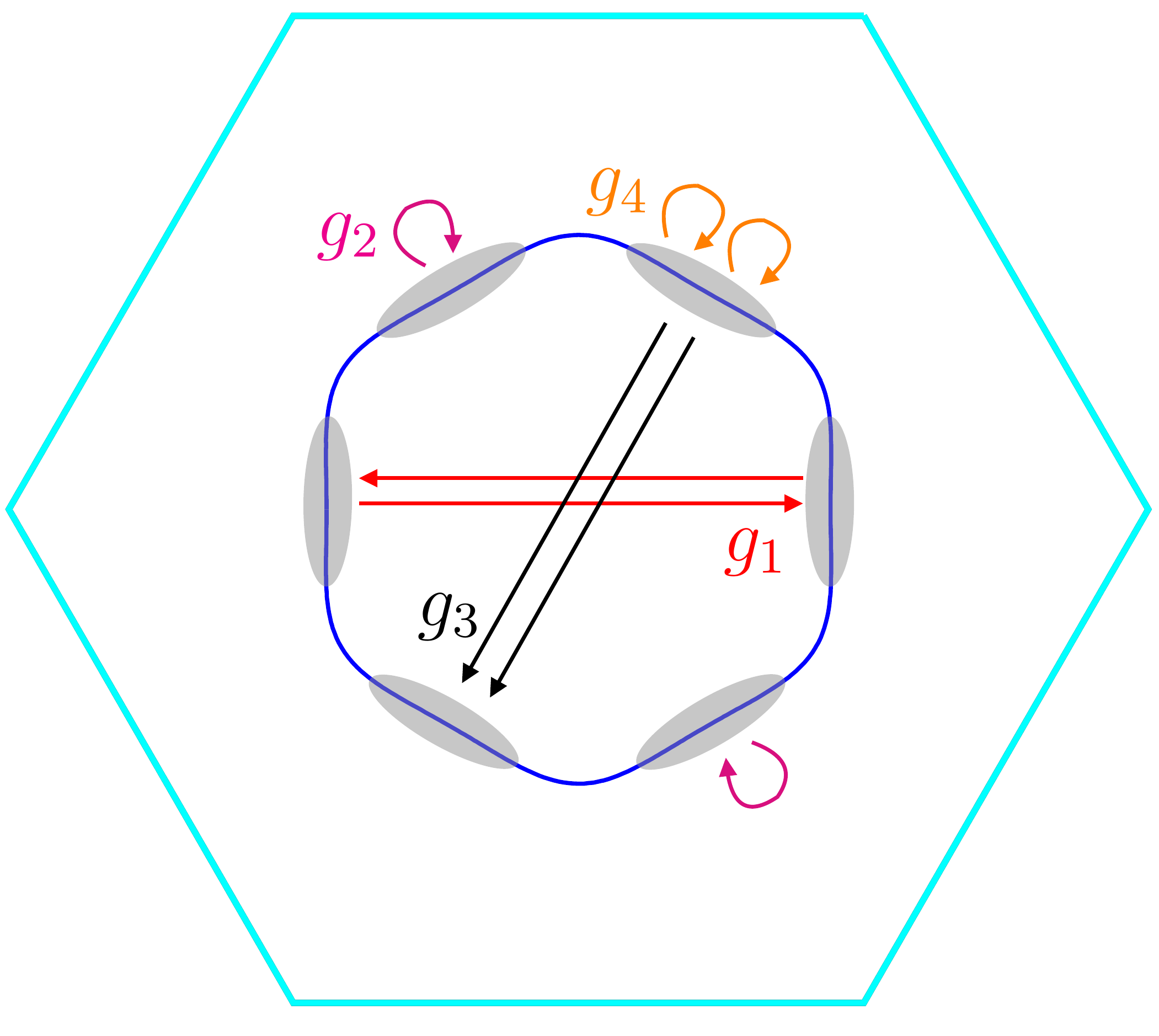}}\hspace{0.25cm}
\subfloat
    [\label{fig:SDW}]
    {\includegraphics[width=0.3\textwidth]{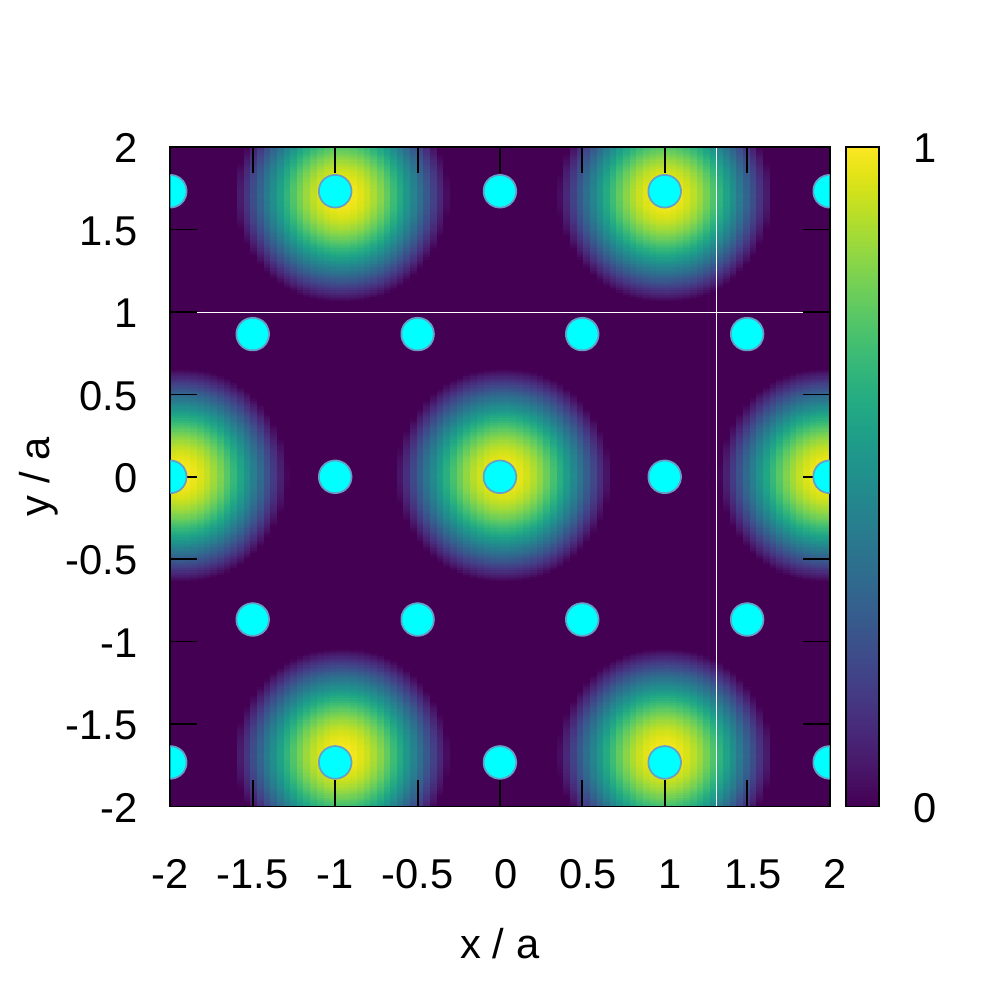}}

\caption{(a) Nested Fermi surface. Three opposite edges are nested with vectors $\bm Q_{1}, \bm Q_{2}, \bm Q_{3}$. We have chosen to label pairs of patches, i.e. $\{1,1'\}$,  which are connected by a $\bm Q_i$ vector. (b) The interactions detailed in the patch model \eqref{equ:patchModelLagrangian}. (c) The $z$-component of the spin density wave (\ref{SDWorder}), with $\bm\varphi_1=\bm\varphi_2=\bm\varphi_3=(0,0,1)$. Light blue circles represent the positions of the  TAL sites. The nesting vectors, $\bm Q_i$, are determined numerically. The corresponding SDW lattice constant is, from $\bm Q_i\cdot \bm L_i=2\pi$, found to be $|L_i|\approx 2 a$.}
\label{fig:nestingSDW}
\end{figure*}
The flat band together with the second Dirac point (e.g. Fig. \ref{fig:AGbandstructure_W=2p5}) is relevant to the effective kagome model but we can also consider the flat band in isolation. In TALs the flat band still has some small dispersion. As mentioned above, the two minima of the flat band are approximately quadratic (Fig. \ref{fig:AG_FB}) and we can assign to them an effective mass $m^{*}$ (Fig. \ref{fig:effMass}). If we suppose that the chemical potential is within the quadratic part of the flat band (which amounts to roughly less than one third filling of that band) then we can ask whether the Stoner criterion for ferromagnetism\cite{khomskiiBasicAspectsCondensed2010} is satisfied. In our system the Stoner criterion can be written as,
\begin{align}\label{equ:stoner}
    S \equiv \frac{A_{cell}}{3} U_{0} | \Pi_{0} | > 1
\end{align}
Where $\Pi_{0}$ is the 2D polarisation operator, which contains a factor 2 due there being two minima in the flat band, at the $\bm K, \bm K'$ points. The factor $A_{cell}/3$ is required by the normalisation of our wavefunctions. To normalise the wavefunctions we need to obtain unity after summation over the \textbf{three} sublattices of the kagome lattice and this introduces a factor $1/3$ to the normalisation coefficient. In two dimensions the polarisation operator is,
\begin{align*}
    \Pi_{0} = - \frac{m^{*}}{\pi \hbar^{2}}
\end{align*}
We have plotted $m^{*}(W)$ and $S(W)$ in figure \ref{fig:effMassStoner}. Within both the electron and hole flatbands the Stoner parameter, $S$, takes values  larger than 1, indicating that the lower part of the flatband is well within the Stoner regime. Such large values follow from $S$ being proportional to the density of states, which diverges within the flat band. In connection to this point, the significant difference in $S$ between holes and electrons is due to the hole flatband width being several times smaller.

The presence of a ferromagnetic phase in a 2D material can be determined via observation of an anomalous Hall resistivity. The anomalous Hall effect is exhibited in systems with spin-orbit coupling and manifests as a hysteresis in $R_{xy}$, measured for up and down sweeps of an external magnetic field (see, for example, Ref. \cite{matsuoka_spinorbit-induced_2021}). Since this effect relies on spin-orbit coupling it is only observable for holes and not electrons.

\subsection{Incommensurate Spin Density Wave}

Suppose we tune the system below the ferromagnetic Stoner instability, $S < 1$, such that there is no ferromagnetism. This is achieved by lowering $W / E_{0}$. At $W = 0.25 E_{0}$ (Fig. \ref{fig:holeBands_W=0p25}), for example, we find that a single ``antiparticle" Fermi surface is formed which centers around $\Gamma$ when the chemical potential is near the top of the third energy band (i.e. the flatband). We find that there exists a critical chemical potential, $\mu = \mu_c$, such that the Fermi surface exhibits nesting, with nesting vectors $\bm Q_i$. This Fermi surface and the nesting vectors are shown in figure \ref{fig:nesting}. A nested Fermi surface with nesting vector $\bm{Q}$ typically promotes spin and/or charge density wave (SDW/CDW) ordering with wavevector $\bm Q$, due to logarithmic enhancement of the polarization operator (and hence of the corresponding Stoner parameter).

\begin{figure*}[!t]
    \centering
    {\includegraphics[width=0.50\textwidth]{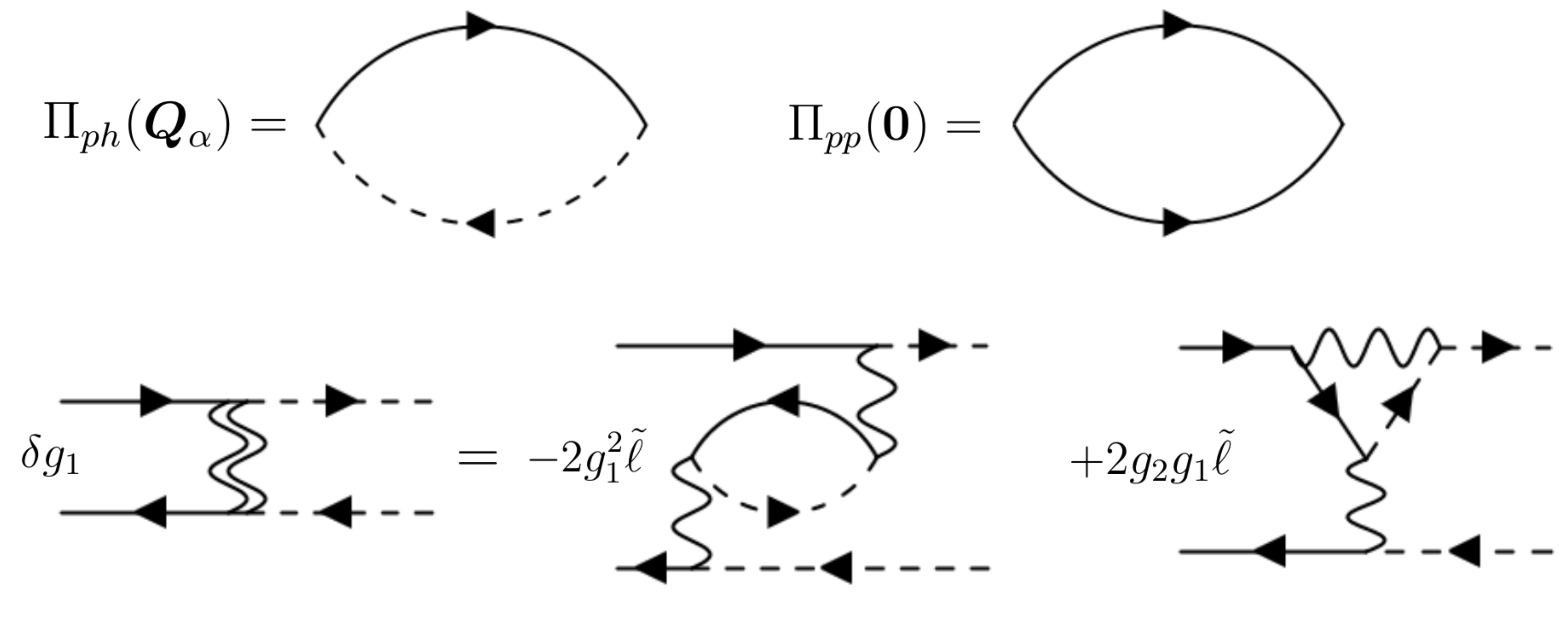}}
\renewcommand{\thesubfigure}{b}
\subfloat
    [\label{fig:RG1}]
    {\includegraphics[width=0.25\textwidth]{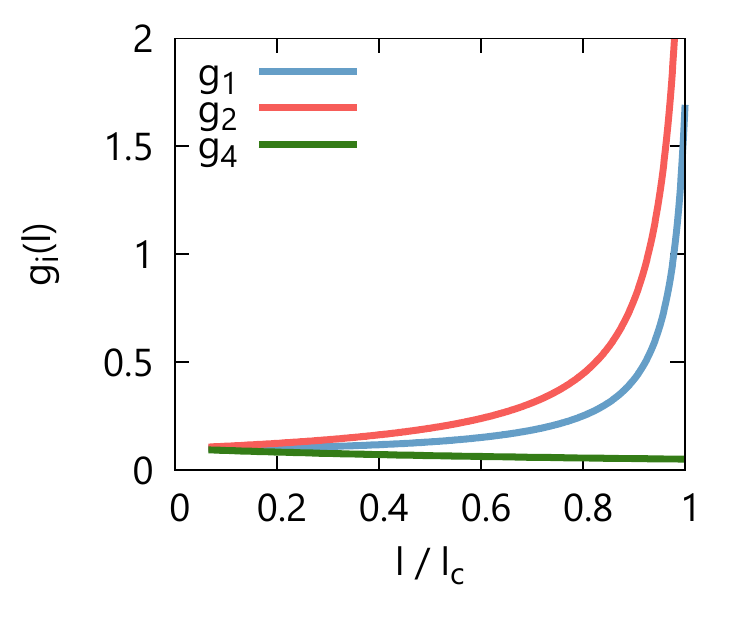}}

\renewcommand{\thesubfigure}{a}
\subfloat
    [\label{fig:RG_diagrams2}]
    {\includegraphics[width=0.50\textwidth]{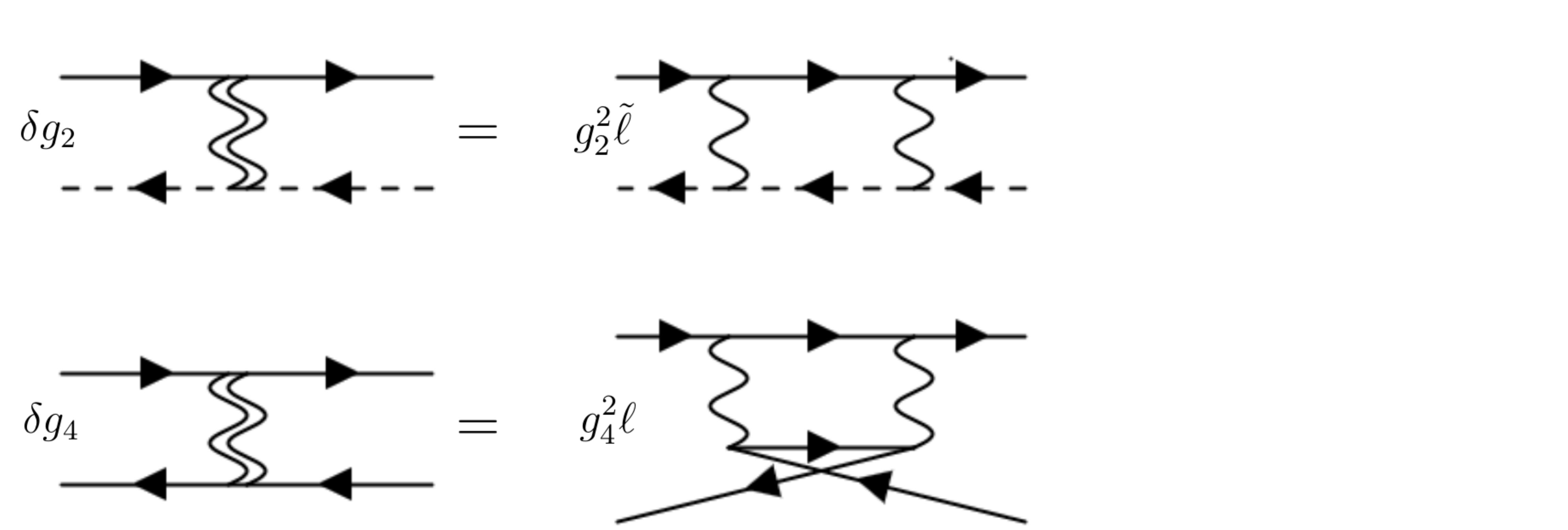}}
\renewcommand{\thesubfigure}{c}
\subfloat[c]
    [\label{fig:RG2}]
    {\includegraphics[width=0.25\textwidth]{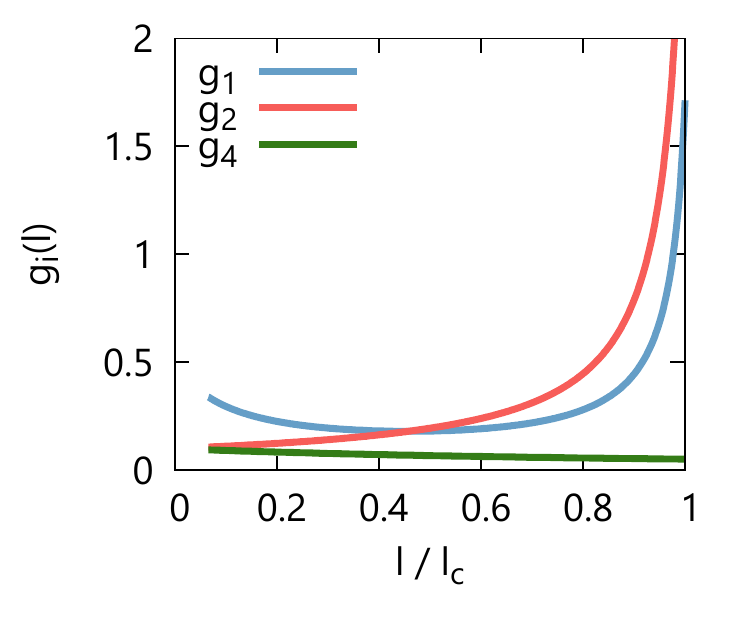}}

\caption{(a) The two relevant Fermion bubble diagrams and coupling corrections $\delta g_{1}$, $\delta g_{2}$ and $\delta g_{4}$. Solid lines denote Fermion propagators on a given patch $\alpha$, while dotted lines denote a Fermion propagator on a different patch $\beta \neq \alpha$. The single wavy line corresponds to a bare $g_{i}$ and the double wavy line refers to a renormalised $g_{i}$. The particle-hole bubble diagram is at nesting momentum $\bm{Q}_{\alpha}$, and the particle-particle diagram at zero momentum. The quantities, $\delta g_{1}$, $\delta g_{2}$ and $\delta g_{4}$, are the leading logarithmic (i.e. first order in $\ell$, $\tilde{\ell}$) corrections to the bare couplings $g_{1}$, $g_{2}$ and $g_{4}$. Since wavy lines can be shrunk to a point, all internal integrations of propagators in (a) correspond to either $\Pi_{ph}(\bm Q_\alpha) = \tilde{\ell}$ or $\Pi_{pp}(\bm 0) = \ell$, as shown. (b and c) Plot of $g_{1}(\ell)$, $g_{2}(\ell)$, and $g_{4}(\ell)$ in blue, red and green, respectively. (b) and (c) correspond to two different sets of initial values $g_{1}(\ell=0)$,  $g_{2}(\ell=0)$, and $g_{4}(\ell=0)$. Both (b) and (c) correspond to $d_{1} = 1$. In both cases, $g_{2}(\ell)$ diverges more strongly than the other $g_{i}$.}
\label{fig:RG}
\end{figure*}

To determine which order (SDW, CDW or SC) is promoted by nesting, we appeal to the following patch model \cite{chubukovChiralSuperconductivity2012, Maiti2013}, which is a minimal model to account for interactions on a nested Fermi surface,

\begin{align}\label{equ:patchModelLagrangian}
    {\cal L}_\alpha^{(0)} =&
    \frac{1}{2} \sum_{p, \alpha, \sigma}
    \begin{bmatrix}
        \psi^\dag_{\alpha,\sigma,p} \\
        \psi^\dag_{\alpha',\sigma p}
    \end{bmatrix}
    \begin{bmatrix}
        \omega - \varepsilon_{\alpha,p} && 0 \\
        0 && \omega + \varepsilon_{\alpha,p}
    \end{bmatrix}
    \begin{bmatrix}
        \psi_{\alpha,\sigma,p} \\
        \psi_{\alpha',\sigma, p}
    \end{bmatrix} \nonumber \\
    {\cal L}_\alpha^{(int)} =& -
    \frac{1}{2} \sum_{ p_{i}, \alpha \neq \beta, \sigma }
    \Big[
        g_{1} \psi^{\dag}_{\alpha, \sigma, p_{1}}
              \psi^\dag_{\beta, \sigma', p_{2}}
              \psi_{\alpha, \sigma', p_{3}}
              \psi_{\beta, \sigma, p_{4}} \nonumber \\
    & \hspace{1.7cm} +
        g_2 \psi^\dag_{\alpha,\sigma,p_1}
            \psi^\dag_{\beta,\sigma',p_2}
            \psi_{\beta,\sigma',p_3}
            \psi_{\alpha,\sigma,p_4} \nonumber \\
    & \hspace{1.7cm} +
        g_3 \psi^\dag_{\alpha,\sigma,p_1}
            \psi^\dag_{\alpha,\sigma',p_2}
            \psi_{\beta,\sigma',p_3}
            \psi_{\beta,\sigma,p_4}
    \Big] \nonumber \\
    & - \frac{1}{2} \sum_{p_i,\sigma}
        g_4 \psi^\dag_{\alpha,\sigma,p_1}
            \psi^\dag_{\alpha,\sigma',p_2}
            \psi_{\alpha,\sigma',p_3}
        \psi_{\alpha,\sigma,p_4}
\end{align}

The model ${\cal L}_\alpha^{(0)}$ describes the Fermi surface patch $\alpha$, and its nesting with patch $\alpha'$ (see Fig. \ref{fig:nesting}) with all patches treated as being independent of each other. Here, $p$ represents the momentum of patch $\alpha$ and $\sigma = -\sigma'$ labels the spin. Our notation in ${\cal L}_\alpha^{(0)}$ takes into account the nesting property $\varepsilon_{\alpha',p}=-\varepsilon_{\alpha,p}$.

%The model ${\cal L}_\alpha^{(0)}$ describes the Fermi surface patch $\alpha$, and it's nesting with patch $\alpha'$, which is connected by $\bm Q_\alpha$, i.e. patches $1$ and $1'$ are connected by $\bm{Q}_{1}$ (Fig. \ref{fig:nesting}). In ${\cal L}_\alpha^{(0)}$ we have already accounted for the nesting property $\varepsilon_{\alpha',p}=-\varepsilon_{\alpha,p}$, where $p$ is the momentum belonging to patch $\alpha$. Everywhere $\sigma'= -\sigma$, labels the spin.

Interactions between patches are accounted for in the second term of Eqn. \ref{equ:patchModelLagrangian}, ${\cal L}_\alpha^{(int)}$. Here we account for allowed four-Fermion interaction processes both for Fermions in different patches and within the same patch. Note that momentum conservation is assumed, $\bm{p}_1 + \bm{p}_2 = \bm{p}_3 + \bm{p}_4$ (modulo a reciprocal lattice vector $\bm G_i$), within the summation. The couplings $g_{i}$ correspond to the following processes: $g_{1}$ is a patch-exchange interaction; $g_{2}$ is a density-density interaction for fermions on different patches; $g_{3}$ is pair hoping between patches; $g_{4}$ is a density-density interaction for fermions on the same patch. Such interaction processes are represented in Fig. \ref{fig:nestingSDW}(b). The interaction process $g_{3}$ only exists for special momentum transfer $2\bm Q_{\alpha} = \bm G_{i}$, and is essential to generate superconductivity\cite{chubukovChiralSuperconductivity2012} (we discuss this in following section, \ref{superconductivity}). For the doped flat band, Fermi surface nesting occurs with vectors $2 \bm{Q}_{\alpha} \neq \bm{G}_{i}$, so we must set $g_3 = 0$ since it does not conserve momentum (modulo $\bm{G}_{i}$).

The interaction vertices $g_{i} \propto U_{0}$, however we leave them unevaluated and treat them as the momentum independent parameters. Upon renormalisation, which is described below, the interaction parameters $g_i$ gain a logarithmic scale dependence, the divergence of which ultimately determines the nature of the ordered state, i.e. CDW, SDW or SC.

To setup the renormalisation procedure, we consider the Fermion bubble operators for particle-particle (at zero momentum transfer) and particle-hole (at $Q_{\alpha}$ momentum transfer), shown in Fig. \ref{fig:RG_diagrams2}, which are given by,

\begin{align*}
    \Pi_{pp}(\bm 0) &=
    \frac{1}{4} \nu_0 \ln\frac{\Lambda}{T} \equiv \ell \nonumber \\
    \Pi_{ph}(\bm Q_{\alpha}) &=
    \frac{1}{4} \nu_0 \ln\frac{\Lambda}{\max\{T,\mu\}} \equiv \tilde{\ell}
\end{align*}

where $\nu_{0}$ is the single-spin density of states. For simplicity, we will henceforth set $\tilde{\ell} = \ell$.

Given these logarithmic bubble operators, we perform a leading logarithm resummation of diagrams which renormalise the bare coupling constants, $g_i$. This procedure follows from previous works  \cite{chubukovChiralSuperconductivity2012, Schulz1987, Dzyaloshinskii1987}. The resulting RG equations for the coupling constants follow from the diagrammatic series of Fig. \ref{fig:RG_diagrams2},

\begin{align*}
    \frac{d g_1}{d\ell} &= 2 g_1(g_2-g_1), \\
    \frac{d g_2}{d\ell} &=   g_2^2, \\
    \frac{d g_4}{d\ell} &= - g_4^2
\end{align*}

Where we assume that $g_i(\Lambda) = g_i(\ell=0) > 0$. For various initial values, we find that $g_2(\ell) > g_1(\ell)$ in the limit $\ell \to \ell_c$ (see Figs. \ref{fig:RG1} and \ref{fig:RG2}). This implies SDW state is the leading instability.

Performing a Landau-Ginzburg type expansion of the free energy, we obtain the following structure for the real space SDW order parameter,

\begin{align}\label{SDWorder}
    \bm{\varphi}(\bm r) &=
    \bm{\varphi}_1 \cos(\bm Q_{1}\cdot \bm r) +
    \bm{\varphi}_2 \cos(\bm Q_{2}\cdot \bm r) +
    \bm{\varphi}_3 \cos(\bm Q_{3}\cdot \bm r)
\end{align}

Here $\bm \varphi_i \in {\mathbb R}^3$ are constant real vectors with equal magnitude, but arbitrary orientation. Hence this order parameter is highly degenerate. We plot one such realisation, $\bm\varphi_1 = \bm\varphi_2 = \bm\varphi_3 = (0,0,1)$, in Figure \ref{fig:SDW}.

This order does not fully gap the charge carriers at the Fermi surface, and so would not be straightforward to detect in transport measurements. However, the SDW ordering wavevectors, $\bm Q_i$, reconstruct the Fermi surface, to form small fermi pockets; such an effect could be measured via, e.g., Shubnikov-de Haas oscillations.

\subsection{Superconductivity}\label{superconductivity}

It is worthwhile to mention the possibility of nesting-induced superconductivity, which follows from the formalism developed by Chubukov et. al.\cite{chubukovChiralSuperconductivity2012} and adapted in the previous section. That work predicts that graphene, doped such that the Fermi surface passes through the $M$ point of the Brillouin zone (Fig. \ref{fig:BZ}), will exhibit superconductivity. The Fermi surface at this point is hexagonal with vertices which touch the $M$-points. Such a Fermi surface exhibits nesting and is also found in the band structure of our artificial lattice when the chemical potential is just above or just below the Dirac cones (see bands 1, 2 and 4, 5 in figure \ref{fig:AGbandstructure_W=2p5}, for example). In this case the nesting vector is equal to a reciprocal lattice vector ($2 Q_{\alpha \beta} = G_{i}$) and hence the coupling, $g_3$, is allowed. Inclusion of this coupling dramatically influences the RG flow, promoting $d + id$ superconductivity as the leading instability. For the band structures we consider in this work (e.g Fig. \ref{fig:AGbandstructure_W=2p5} and Fig. \ref{fig:holeBands_W=1p00}), this situation is realised at four different values of $E_{F}$, each which correspond to a van-Hove singularity in the density of states.

In addition to nesting-induced superconductivity for doping at the $M$-point, there is a recent proposal\cite{LiInghamScammell2020} for pseudospin superconductivity in Dirac-like bands which occurs for doping slightly above or below the $K$ and $K'$ points. This situation is directly relevant to the two sets of Dirac bands realised here.

\section{Conclusions}

Previously, two-dimensional artificial triangular anti-dot lattices in a semiconductor have attracted attention due to the possibility of studying Dirac and topological physics. In the present work we shift the focus to electron-electron correlation effects. To be specific we concentrate on GaAs and come to the following conclusions.

(i) At a sufficiently strong anti-dot potential modulation, $W$, the 3rd, 4th and 5th minibands of the system mimic the dispersion of an emergent kagome lattice. Hence, we map the triangular anti-dot system to an effective Hubbard model on a kagome lattice. The model manifests strong electron-electron correlations.

(ii) We demonstrate the significant practical experimental advantage of holes compared to electrons. Due to the spin-orbit interaction of holes, the strength of the potential modulation, $W$, necessary to access the strongly correlated regime is 2 to 3 times smaller for holes than for electrons.

(iii) We demonstrate that, at a sufficiently large $W$ and dependent on the value of the chemical potential, the system develops a Mott transition and also several kinds of commensurate charge density waves.

(iv) We also show that, in the regime of not too large $W$ (precursor to strong correlations) and dependent on the value of the chemical potential, the system has a robust ferromagnetic Stoner instability and also superconducting, charge density wave and spin density wave instabilities.

\begin{acknowledgements}

We wish to acknowledge useful discussions with A. R. Hamilton, J. Ingham, O. Klochan, T. Li and D. Wang. This research was supported by an Australian Government Research Training Program (RTP) Scholarship. We have also received support from the Australian Research Council Centre of Excellence in Future Low-Energy Electronics Technology (FLEET) (CE170100039).

\end{acknowledgements}

\bibliography{bibliography.bib}

\begin{thebibliography}{27}
\expandafter\ifx\csname natexlab\endcsname\relax\def\natexlab#1{#1}\fi
\expandafter\ifx\csname bibnamefont\endcsname\relax
  \def\bibnamefont#1{#1}\fi
\expandafter\ifx\csname bibfnamefont\endcsname\relax
  \def\bibfnamefont#1{#1}\fi
\expandafter\ifx\csname citenamefont\endcsname\relax
  \def\citenamefont#1{#1}\fi
\expandafter\ifx\csname url\endcsname\relax
  \def\url#1{\texttt{#1}}\fi
\expandafter\ifx\csname urlprefix\endcsname\relax\def\urlprefix{URL }\fi
\providecommand{\bibinfo}[2]{#2}
\providecommand{\eprint}[2][]{\url{#2}}

\bibitem[{\citenamefont{Balents et~al.}(2020)\citenamefont{Balents, Dean,
  Efetov, and Young}}]{balents_superconductivity_2020}
\bibinfo{author}{\bibfnamefont{L.}~\bibnamefont{Balents}},
  \bibinfo{author}{\bibfnamefont{C.~R.} \bibnamefont{Dean}},
  \bibinfo{author}{\bibfnamefont{D.~K.} \bibnamefont{Efetov}},
  \bibnamefont{and} \bibinfo{author}{\bibfnamefont{A.~F.} \bibnamefont{Young}},
  \bibinfo{journal}{Nature Physics} \textbf{\bibinfo{volume}{16}},
  \bibinfo{pages}{725} (\bibinfo{year}{2020}), ISSN \bibinfo{issn}{1745-2481},
  \bibinfo{note}{number: 7 Publisher: Nature Publishing Group},
  \urlprefix\url{https://www.nature.com/articles/s41567-020-0906-9}.

\bibitem[{\citenamefont{Zhang et~al.}(2020)\citenamefont{Zhang, Wang, Watanabe,
  Taniguchi, Ueno, Tutuc, and LeRoy}}]{zhang_flat_2020}
\bibinfo{author}{\bibfnamefont{Z.}~\bibnamefont{Zhang}},
  \bibinfo{author}{\bibfnamefont{Y.}~\bibnamefont{Wang}},
  \bibinfo{author}{\bibfnamefont{K.}~\bibnamefont{Watanabe}},
  \bibinfo{author}{\bibfnamefont{T.}~\bibnamefont{Taniguchi}},
  \bibinfo{author}{\bibfnamefont{K.}~\bibnamefont{Ueno}},
  \bibinfo{author}{\bibfnamefont{E.}~\bibnamefont{Tutuc}}, \bibnamefont{and}
  \bibinfo{author}{\bibfnamefont{B.~J.} \bibnamefont{LeRoy}},
  \bibinfo{journal}{Nature Physics} \textbf{\bibinfo{volume}{16}},
  \bibinfo{pages}{1093} (\bibinfo{year}{2020}), ISSN \bibinfo{issn}{1745-2481},
  \bibinfo{note}{number: 11 Publisher: Nature Publishing Group},
  \urlprefix\url{https://www.nature.com/articles/s41567-020-0958-x}.

\bibitem[{\citenamefont{Kang et~al.}(2020)\citenamefont{Kang, Ye, Fang, You,
  Levitan, Han, Facio, Jozwiak, Bostwick, Rotenberg et~al.}}]{kang_dirac_2020}
\bibinfo{author}{\bibfnamefont{M.}~\bibnamefont{Kang}},
  \bibinfo{author}{\bibfnamefont{L.}~\bibnamefont{Ye}},
  \bibinfo{author}{\bibfnamefont{S.}~\bibnamefont{Fang}},
  \bibinfo{author}{\bibfnamefont{J.-S.} \bibnamefont{You}},
  \bibinfo{author}{\bibfnamefont{A.}~\bibnamefont{Levitan}},
  \bibinfo{author}{\bibfnamefont{M.}~\bibnamefont{Han}},
  \bibinfo{author}{\bibfnamefont{J.~I.} \bibnamefont{Facio}},
  \bibinfo{author}{\bibfnamefont{C.}~\bibnamefont{Jozwiak}},
  \bibinfo{author}{\bibfnamefont{A.}~\bibnamefont{Bostwick}},
  \bibinfo{author}{\bibfnamefont{E.}~\bibnamefont{Rotenberg}},
  \bibnamefont{et~al.}, \bibinfo{journal}{Nature Materials}
  \textbf{\bibinfo{volume}{19}}, \bibinfo{pages}{163} (\bibinfo{year}{2020}),
  ISSN \bibinfo{issn}{1476-4660}, \bibinfo{note}{number: 2 Publisher: Nature
  Publishing Group},
  \urlprefix\url{https://www.nature.com/articles/s41563-019-0531-0}.

\bibitem[{\citenamefont{Cao et~al.}(2018)\citenamefont{Cao, Fatemi, Fang,
  Watanabe, Taniguchi, Kaxiras, and Jarillo-Herrero}}]{cao_unconventional_2018}
\bibinfo{author}{\bibfnamefont{Y.}~\bibnamefont{Cao}},
  \bibinfo{author}{\bibfnamefont{V.}~\bibnamefont{Fatemi}},
  \bibinfo{author}{\bibfnamefont{S.}~\bibnamefont{Fang}},
  \bibinfo{author}{\bibfnamefont{K.}~\bibnamefont{Watanabe}},
  \bibinfo{author}{\bibfnamefont{T.}~\bibnamefont{Taniguchi}},
  \bibinfo{author}{\bibfnamefont{E.}~\bibnamefont{Kaxiras}}, \bibnamefont{and}
  \bibinfo{author}{\bibfnamefont{P.}~\bibnamefont{Jarillo-Herrero}},
  \bibinfo{journal}{Nature} \textbf{\bibinfo{volume}{556}}, \bibinfo{pages}{43}
  (\bibinfo{year}{2018}), ISSN \bibinfo{issn}{1476-4687},
  \bibinfo{note}{number: 7699 Publisher: Nature Publishing Group},
  \urlprefix\url{https://www.nature.com/articles/nature26160}.

\bibitem[{\citenamefont{Liu et~al.}(2014)\citenamefont{Liu, Liu, and
  Wu}}]{liu_exotic_2014}
\bibinfo{author}{\bibfnamefont{Z.}~\bibnamefont{Liu}},
  \bibinfo{author}{\bibfnamefont{F.}~\bibnamefont{Liu}}, \bibnamefont{and}
  \bibinfo{author}{\bibfnamefont{Y.-S.} \bibnamefont{Wu}},
  \bibinfo{journal}{Chinese Physics B} \textbf{\bibinfo{volume}{23}},
  \bibinfo{pages}{077308} (\bibinfo{year}{2014}), ISSN
  \bibinfo{issn}{1674-1056}, \bibinfo{note}{publisher: IOP Publishing}.

\bibitem[{\citenamefont{Ortiz et~al.}(2020)\citenamefont{Ortiz, Teicher, Hu,
  Zuo, Sarte, Schueller, Abeykoon, Krogstad, Rosenkranz, Osborn
  et~al.}}]{Ortiz2020}
\bibinfo{author}{\bibfnamefont{B.~R.} \bibnamefont{Ortiz}},
  \bibinfo{author}{\bibfnamefont{S.~M.~L.} \bibnamefont{Teicher}},
  \bibinfo{author}{\bibfnamefont{Y.}~\bibnamefont{Hu}},
  \bibinfo{author}{\bibfnamefont{J.~L.} \bibnamefont{Zuo}},
  \bibinfo{author}{\bibfnamefont{P.~M.} \bibnamefont{Sarte}},
  \bibinfo{author}{\bibfnamefont{E.~C.} \bibnamefont{Schueller}},
  \bibinfo{author}{\bibfnamefont{A.~M.~M.} \bibnamefont{Abeykoon}},
  \bibinfo{author}{\bibfnamefont{M.~J.} \bibnamefont{Krogstad}},
  \bibinfo{author}{\bibfnamefont{S.}~\bibnamefont{Rosenkranz}},
  \bibinfo{author}{\bibfnamefont{R.}~\bibnamefont{Osborn}},
  \bibnamefont{et~al.}, \bibinfo{journal}{Phys. Rev. Lett.}
  \textbf{\bibinfo{volume}{125}}, \bibinfo{pages}{247002}
  (\bibinfo{year}{2020}),
  \urlprefix\url{https://link.aps.org/doi/10.1103/PhysRevLett.125.247002}.

\bibitem[{\citenamefont{Zhu et~al.}(2021)\citenamefont{Zhu, Yang, Xia, Yin,
  Wang, Zhao, Dai, Tu, Song, Tao et~al.}}]{Zhu2021}
\bibinfo{author}{\bibfnamefont{C.~C.} \bibnamefont{Zhu}},
  \bibinfo{author}{\bibfnamefont{X.~F.} \bibnamefont{Yang}},
  \bibinfo{author}{\bibfnamefont{W.}~\bibnamefont{Xia}},
  \bibinfo{author}{\bibfnamefont{Q.~W.} \bibnamefont{Yin}},
  \bibinfo{author}{\bibfnamefont{L.~S.} \bibnamefont{Wang}},
  \bibinfo{author}{\bibfnamefont{C.~C.} \bibnamefont{Zhao}},
  \bibinfo{author}{\bibfnamefont{D.~Z.} \bibnamefont{Dai}},
  \bibinfo{author}{\bibfnamefont{C.~P.} \bibnamefont{Tu}},
  \bibinfo{author}{\bibfnamefont{B.~Q.} \bibnamefont{Song}},
  \bibinfo{author}{\bibfnamefont{Z.~C.} \bibnamefont{Tao}},
  \bibnamefont{et~al.}, \emph{\bibinfo{title}{Double-dome superconductivity
  under pressure in the v-based kagome metals av3sb5 (a = rb and k)}}
  (\bibinfo{year}{2021}), \eprint{2104.14487}.

\bibitem[{\citenamefont{Chen et~al.}(2021)\citenamefont{Chen, Wang, Yin, Gu,
  Jiang, Tu, Gong, Uwatoko, Sun, Lei et~al.}}]{Chen2021}
\bibinfo{author}{\bibfnamefont{K.~Y.} \bibnamefont{Chen}},
  \bibinfo{author}{\bibfnamefont{N.~N.} \bibnamefont{Wang}},
  \bibinfo{author}{\bibfnamefont{Q.~W.} \bibnamefont{Yin}},
  \bibinfo{author}{\bibfnamefont{Y.~H.} \bibnamefont{Gu}},
  \bibinfo{author}{\bibfnamefont{K.}~\bibnamefont{Jiang}},
  \bibinfo{author}{\bibfnamefont{Z.~J.} \bibnamefont{Tu}},
  \bibinfo{author}{\bibfnamefont{C.~S.} \bibnamefont{Gong}},
  \bibinfo{author}{\bibfnamefont{Y.}~\bibnamefont{Uwatoko}},
  \bibinfo{author}{\bibfnamefont{J.~P.} \bibnamefont{Sun}},
  \bibinfo{author}{\bibfnamefont{H.~C.} \bibnamefont{Lei}},
  \bibnamefont{et~al.}, \bibinfo{journal}{Phys. Rev. Lett.}
  \textbf{\bibinfo{volume}{126}}, \bibinfo{pages}{247001}
  (\bibinfo{year}{2021}),
  \urlprefix\url{https://link.aps.org/doi/10.1103/PhysRevLett.126.247001}.

\bibitem[{\citenamefont{Ortiz et~al.}(2021)\citenamefont{Ortiz, Sarte, Kenney,
  Graf, Teicher, Seshadri, and Wilson}}]{Ortiz2021}
\bibinfo{author}{\bibfnamefont{B.~R.} \bibnamefont{Ortiz}},
  \bibinfo{author}{\bibfnamefont{P.~M.} \bibnamefont{Sarte}},
  \bibinfo{author}{\bibfnamefont{E.~M.} \bibnamefont{Kenney}},
  \bibinfo{author}{\bibfnamefont{M.~J.} \bibnamefont{Graf}},
  \bibinfo{author}{\bibfnamefont{S.~M.~L.} \bibnamefont{Teicher}},
  \bibinfo{author}{\bibfnamefont{R.}~\bibnamefont{Seshadri}}, \bibnamefont{and}
  \bibinfo{author}{\bibfnamefont{S.~D.} \bibnamefont{Wilson}},
  \bibinfo{journal}{Phys. Rev. Materials} \textbf{\bibinfo{volume}{5}},
  \bibinfo{pages}{034801} (\bibinfo{year}{2021}),
  \urlprefix\url{https://link.aps.org/doi/10.1103/PhysRevMaterials.5.034801}.

\bibitem[{\citenamefont{Ni et~al.}(2021)\citenamefont{Ni, Ma, Zhang, Yuan,
  Yang, Lu, Wang, Sun, Zhao, Li et~al.}}]{Ni2021}
\bibinfo{author}{\bibfnamefont{S.}~\bibnamefont{Ni}},
  \bibinfo{author}{\bibfnamefont{S.}~\bibnamefont{Ma}},
  \bibinfo{author}{\bibfnamefont{Y.}~\bibnamefont{Zhang}},
  \bibinfo{author}{\bibfnamefont{J.}~\bibnamefont{Yuan}},
  \bibinfo{author}{\bibfnamefont{H.}~\bibnamefont{Yang}},
  \bibinfo{author}{\bibfnamefont{Z.}~\bibnamefont{Lu}},
  \bibinfo{author}{\bibfnamefont{N.}~\bibnamefont{Wang}},
  \bibinfo{author}{\bibfnamefont{J.}~\bibnamefont{Sun}},
  \bibinfo{author}{\bibfnamefont{Z.}~\bibnamefont{Zhao}},
  \bibinfo{author}{\bibfnamefont{D.}~\bibnamefont{Li}}, \bibnamefont{et~al.},
  \bibinfo{journal}{Chinese Physics Letters} \textbf{\bibinfo{volume}{38}},
  \bibinfo{pages}{057403} (\bibinfo{year}{2021}),
  \urlprefix\url{https://doi.org/10.1088/0256-307x/38/5/057403}.

\bibitem[{\citenamefont{Jiang et~al.}(2021)\citenamefont{Jiang, Yin, Denner,
  Shumiya, Ortiz, Xu, Guguchia, He, Hossain, Liu et~al.}}]{Jiang2021}
\bibinfo{author}{\bibfnamefont{Y.-X.} \bibnamefont{Jiang}},
  \bibinfo{author}{\bibfnamefont{J.-X.} \bibnamefont{Yin}},
  \bibinfo{author}{\bibfnamefont{M.~M.} \bibnamefont{Denner}},
  \bibinfo{author}{\bibfnamefont{N.}~\bibnamefont{Shumiya}},
  \bibinfo{author}{\bibfnamefont{B.~R.} \bibnamefont{Ortiz}},
  \bibinfo{author}{\bibfnamefont{G.}~\bibnamefont{Xu}},
  \bibinfo{author}{\bibfnamefont{Z.}~\bibnamefont{Guguchia}},
  \bibinfo{author}{\bibfnamefont{J.}~\bibnamefont{He}},
  \bibinfo{author}{\bibfnamefont{M.~S.} \bibnamefont{Hossain}},
  \bibinfo{author}{\bibfnamefont{X.}~\bibnamefont{Liu}}, \bibnamefont{et~al.},
  \bibinfo{journal}{Nature Materials}  (\bibinfo{year}{2021}), ISSN
  \bibinfo{issn}{1476-4660},
  \urlprefix\url{https://doi.org/10.1038/s41563-021-01034-y}.

\bibitem[{\citenamefont{Tkachenko et~al.}(2015)\citenamefont{Tkachenko,
  Tkachenko, Terekhov, and Sushkov}}]{tkachenkoEffectsCoulombScreening2015}
\bibinfo{author}{\bibfnamefont{O.~A.} \bibnamefont{Tkachenko}},
  \bibinfo{author}{\bibfnamefont{V.~A.} \bibnamefont{Tkachenko}},
  \bibinfo{author}{\bibfnamefont{I.~S.} \bibnamefont{Terekhov}},
  \bibnamefont{and} \bibinfo{author}{\bibfnamefont{O.~P.}
  \bibnamefont{Sushkov}}, \bibinfo{journal}{2D Materials}
  \textbf{\bibinfo{volume}{2}}, \bibinfo{pages}{014010} (\bibinfo{year}{2015}),
  ISSN \bibinfo{issn}{2053-1583}.

\bibitem[{\citenamefont{Wang et~al.}(2021)\citenamefont{Wang, Krix, Sushkov,
  Farrer, Ritchie, Hamilton, and Klochan}}]{wangUnpublished2021}
\bibinfo{author}{\bibfnamefont{D.~Q.} \bibnamefont{Wang}},
  \bibinfo{author}{\bibfnamefont{Z.~E.} \bibnamefont{Krix}},
  \bibinfo{author}{\bibfnamefont{O.~P.} \bibnamefont{Sushkov}},
  \bibinfo{author}{\bibfnamefont{I.}~\bibnamefont{Farrer}},
  \bibinfo{author}{\bibfnamefont{D.~A.} \bibnamefont{Ritchie}},
  \bibinfo{author}{\bibfnamefont{A.~R.} \bibnamefont{Hamilton}},
  \bibnamefont{and} \bibinfo{author}{\bibfnamefont{O.}~\bibnamefont{Klochan}},
  \bibinfo{journal}{To be published}  (\bibinfo{year}{2021}).

\bibitem[{\citenamefont{Du et~al.}(2021)\citenamefont{Du, Liu, Wind,
  Pellegrini, West, Fallahi, Pfeiffer, Manfra, and Pinczuk}}]{Pinczuk2021PRL}
\bibinfo{author}{\bibfnamefont{L.}~\bibnamefont{Du}},
  \bibinfo{author}{\bibfnamefont{Z.}~\bibnamefont{Liu}},
  \bibinfo{author}{\bibfnamefont{S.~J.} \bibnamefont{Wind}},
  \bibinfo{author}{\bibfnamefont{V.}~\bibnamefont{Pellegrini}},
  \bibinfo{author}{\bibfnamefont{K.~W.} \bibnamefont{West}},
  \bibinfo{author}{\bibfnamefont{S.}~\bibnamefont{Fallahi}},
  \bibinfo{author}{\bibfnamefont{L.~N.} \bibnamefont{Pfeiffer}},
  \bibinfo{author}{\bibfnamefont{M.~J.} \bibnamefont{Manfra}},
  \bibnamefont{and} \bibinfo{author}{\bibfnamefont{A.}~\bibnamefont{Pinczuk}},
  \bibinfo{journal}{Phys. Rev. Lett.} \textbf{\bibinfo{volume}{126}},
  \bibinfo{pages}{106402} (\bibinfo{year}{2021}),
  \urlprefix\url{https://link.aps.org/doi/10.1103/PhysRevLett.126.106402}.

\bibitem[{\citenamefont{Luttinger}(1956)}]{luttingerQuantumTheoryCyclotron1956}
\bibinfo{author}{\bibfnamefont{J.~M.} \bibnamefont{Luttinger}},
  \bibinfo{journal}{Phys. Rev.} \textbf{\bibinfo{volume}{102}},
  \bibinfo{pages}{1030} (\bibinfo{year}{1956}),
  \urlprefix\url{https://link.aps.org/doi/10.1103/PhysRev.102.1030}.

\bibitem[{\citenamefont{Miserev and
  Sushkov}(2017)}]{miserevDimensionalReductionLuttinger2017}
\bibinfo{author}{\bibfnamefont{D.~S.} \bibnamefont{Miserev}} \bibnamefont{and}
  \bibinfo{author}{\bibfnamefont{O.~P.} \bibnamefont{Sushkov}},
  \bibinfo{journal}{Phys. Rev. B} \textbf{\bibinfo{volume}{95}},
  \bibinfo{pages}{085431} (\bibinfo{year}{2017}),
  \urlprefix\url{https://link.aps.org/doi/10.1103/PhysRevB.95.085431}.

\bibitem[{\citenamefont{Landau and Lifshitz}(1977)}]{landau_Volume_3}
\bibinfo{author}{\bibfnamefont{L.}~\bibnamefont{Landau}} \bibnamefont{and}
  \bibinfo{author}{\bibfnamefont{E.}~\bibnamefont{Lifshitz}}, in
  \emph{\bibinfo{booktitle}{Quantum Mechanics (Third Edition)}}, edited by
  \bibinfo{editor}{\bibfnamefont{L.}~\bibnamefont{Landau}} \bibnamefont{and}
  \bibinfo{editor}{\bibfnamefont{E.}~\bibnamefont{Lifshitz}}
  (\bibinfo{publisher}{Pergamon}, \bibinfo{year}{1977}), pp.
  \bibinfo{pages}{197--224}, \bibinfo{edition}{third edition} ed., ISBN
  \bibinfo{isbn}{978-0-08-020940-1},
  \urlprefix\url{https://www.sciencedirect.com/science/article/pii/B9780080209401500153}.

\bibitem[{\citenamefont{Winkler}(2003)}]{winklerSpinOrbitCoupling2003}
\bibinfo{author}{\bibfnamefont{R.}~\bibnamefont{Winkler}},
  \emph{\bibinfo{title}{Spin-orbit Coupling Effects in Two-Dimensional Electron
  and Hole Systems}}, Springer Tracts in Modern Physics
  (\bibinfo{publisher}{Springer Berlin Heidelberg}, \bibinfo{year}{2003}), ISBN
  \bibinfo{isbn}{9783540366164},
  \urlprefix\url{https://books.google.com.au/books?id=w4B8CwAAQBAJ}.

\bibitem[{\citenamefont{Sushkov and Castro~Neto}(2013)}]{SushkovNeto2013}
\bibinfo{author}{\bibfnamefont{O.~P.} \bibnamefont{Sushkov}} \bibnamefont{and}
  \bibinfo{author}{\bibfnamefont{A.~H.} \bibnamefont{Castro~Neto}},
  \bibinfo{journal}{Phys. Rev. Lett.} \textbf{\bibinfo{volume}{110}},
  \bibinfo{pages}{186601} (\bibinfo{year}{2013}),
  \urlprefix\url{https://link.aps.org/doi/10.1103/PhysRevLett.110.186601}.

\bibitem[{\citenamefont{Scammell and Sushkov}(2019)}]{ScammellSushkov2019}
\bibinfo{author}{\bibfnamefont{H.~D.} \bibnamefont{Scammell}} \bibnamefont{and}
  \bibinfo{author}{\bibfnamefont{O.~P.} \bibnamefont{Sushkov}},
  \bibinfo{journal}{Phys. Rev. B} \textbf{\bibinfo{volume}{99}},
  \bibinfo{pages}{085419} (\bibinfo{year}{2019}),
  \urlprefix\url{https://link.aps.org/doi/10.1103/PhysRevB.99.085419}.

\bibitem[{\citenamefont{Khomskii}(2010)}]{khomskiiBasicAspectsCondensed2010}
\bibinfo{author}{\bibfnamefont{D.~I.} \bibnamefont{Khomskii}},
  \emph{\bibinfo{title}{Instabilities and phase transitions in electronic
  systems}} (\bibinfo{publisher}{Cambridge University Press},
  \bibinfo{year}{2010}), p. \bibinfo{pages}{188–228}.

\bibitem[{\citenamefont{Matsuoka et~al.}(2021)\citenamefont{Matsuoka, Barnes,
  Ieda, Maekawa, Bahramy, Saika, Takeda, Wadati, Wang, Yoshida
  et~al.}}]{matsuoka_spinorbit-induced_2021}
\bibinfo{author}{\bibfnamefont{H.}~\bibnamefont{Matsuoka}},
  \bibinfo{author}{\bibfnamefont{S.~E.} \bibnamefont{Barnes}},
  \bibinfo{author}{\bibfnamefont{J.}~\bibnamefont{Ieda}},
  \bibinfo{author}{\bibfnamefont{S.}~\bibnamefont{Maekawa}},
  \bibinfo{author}{\bibfnamefont{M.~S.} \bibnamefont{Bahramy}},
  \bibinfo{author}{\bibfnamefont{B.~K.} \bibnamefont{Saika}},
  \bibinfo{author}{\bibfnamefont{Y.}~\bibnamefont{Takeda}},
  \bibinfo{author}{\bibfnamefont{H.}~\bibnamefont{Wadati}},
  \bibinfo{author}{\bibfnamefont{Y.}~\bibnamefont{Wang}},
  \bibinfo{author}{\bibfnamefont{S.}~\bibnamefont{Yoshida}},
  \bibnamefont{et~al.}, \bibinfo{journal}{Nano Letters}
  \textbf{\bibinfo{volume}{21}}, \bibinfo{pages}{1807} (\bibinfo{year}{2021}),
  ISSN \bibinfo{issn}{1530-6984}, \bibinfo{note}{publisher: American Chemical
  Society}, \urlprefix\url{https://doi.org/10.1021/acs.nanolett.0c04851}.

\bibitem[{\citenamefont{{Nandkishore} et~al.}(2012)\citenamefont{{Nandkishore},
  {Levitov}, and {Chubukov}}}]{chubukovChiralSuperconductivity2012}
\bibinfo{author}{\bibfnamefont{R.}~\bibnamefont{{Nandkishore}}},
  \bibinfo{author}{\bibfnamefont{L.~S.} \bibnamefont{{Levitov}}},
  \bibnamefont{and} \bibinfo{author}{\bibfnamefont{A.~V.}
  \bibnamefont{{Chubukov}}}, \bibinfo{journal}{Nature Physics}
  \textbf{\bibinfo{volume}{8}}, \bibinfo{pages}{158} (\bibinfo{year}{2012}),
  \eprint{1107.1903}.

\bibitem[{\citenamefont{Maiti and Chubukov}(2013)}]{Maiti2013}
\bibinfo{author}{\bibfnamefont{S.}~\bibnamefont{Maiti}} \bibnamefont{and}
  \bibinfo{author}{\bibfnamefont{A.~V.} \bibnamefont{Chubukov}},
  \bibinfo{journal}{AIP Conference Proceedings}
  \textbf{\bibinfo{volume}{1550}}, \bibinfo{pages}{3} (\bibinfo{year}{2013}),
  \eprint{https://aip.scitation.org/doi/pdf/10.1063/1.4818400},
  \urlprefix\url{https://aip.scitation.org/doi/abs/10.1063/1.4818400}.

\bibitem[{\citenamefont{Schulz}(1987)}]{Schulz1987}
\bibinfo{author}{\bibfnamefont{H.~J.} \bibnamefont{Schulz}},
  \bibinfo{journal}{Europhysics Letters ({EPL})} \textbf{\bibinfo{volume}{4}},
  \bibinfo{pages}{609} (\bibinfo{year}{1987}),
  \urlprefix\url{https://doi.org/10.1209/0295-5075/4/5/016}.

\bibitem[{\citenamefont{Dzyaloshinskii}(1987)}]{Dzyaloshinskii1987}
\bibinfo{author}{\bibfnamefont{I.}~\bibnamefont{Dzyaloshinskii}},
  \bibinfo{journal}{Sov. Phys. JETP} \textbf{\bibinfo{volume}{66}},
  \bibinfo{pages}{848} (\bibinfo{year}{1987}).

\bibitem[{\citenamefont{Li et~al.}(2020)\citenamefont{Li, Ingham, and
  Scammell}}]{LiInghamScammell2020}
\bibinfo{author}{\bibfnamefont{T.}~\bibnamefont{Li}},
  \bibinfo{author}{\bibfnamefont{J.}~\bibnamefont{Ingham}}, \bibnamefont{and}
  \bibinfo{author}{\bibfnamefont{H.~D.} \bibnamefont{Scammell}},
  \bibinfo{journal}{Phys. Rev. Research} \textbf{\bibinfo{volume}{2}},
  \bibinfo{pages}{043155} (\bibinfo{year}{2020}),
  \urlprefix\url{https://link.aps.org/doi/10.1103/PhysRevResearch.2.043155}.

\end{thebibliography}
\end{document}